\documentclass{aastex631}

\begin{document}

\title{C3NN: Cosmological Correlator Convolutional Neural Network \vspace{1mm} \\
\small an interpretable machine learning tool for cosmological analyses}

\author[0009-0002-7361-4073]{Zhengyangguang Gong}
\affiliation{Universit\"{a}ts-Sternwarte, Fakult\"{a}t f\"{u}r Physik, Ludwig-Maximilians-Universit\"{a}t M\"{u}nchen,\\Scheinerstra{\ss}e 1, 81679 M\"{u}nchen, Germany}
\affiliation{Max Planck Institute for Extraterrestrial Physics, Giessenbachstra{\ss}e 1, 85748 Garching, Germany}

\author[0000-0002-0352-9351]{Anik Halder}
\affiliation{Universit\"{a}ts-Sternwarte, Fakult\"{a}t f\"{u}r Physik, Ludwig-Maximilians-Universit\"{a}t M\"{u}nchen,\\Scheinerstra{\ss}e 1, 81679 M\"{u}nchen, Germany}
\affiliation{Max Planck Institute for Extraterrestrial Physics, Giessenbachstra{\ss}e 1, 85748 Garching, Germany}

\author{Annabelle Bohrdt}
\affiliation{University of Regensburg, Universit\"{a}tsstra{\ss}e 31, Regensburg D-93053, Germany}
\affiliation{Munich Center for Quantum Science and Technology, Schellingstra{\ss}e 4, Munich D-80799, Germany}

\author{Stella Seitz}
\affiliation{Universit\"{a}ts-Sternwarte, Fakult\"{a}t f\"{u}r Physik, Ludwig-Maximilians-Universit\"{a}t M\"{u}nchen,\\Scheinerstra{\ss}e 1, 81679 M\"{u}nchen, Germany}
\affiliation{Max Planck Institute for Extraterrestrial Physics, Giessenbachstra{\ss}e 1, 85748 Garching, Germany}

\author{David Gebauer}
\affiliation{Universit\"{a}ts-Sternwarte, Fakult\"{a}t f\"{u}r Physik, Ludwig-Maximilians-Universit\"{a}t M\"{u}nchen,\\Scheinerstra{\ss}e 1, 81679 M\"{u}nchen, Germany}
 
\begin{abstract}
\noindent Modern cosmological research in large scale structure has witnessed an increasing number of applications of machine learning methods. Among them, Convolutional Neural Networks (CNNs) have received substantial attention due to their outstanding performance in image classification, cosmological parameter inference and various other tasks. However, many models which make use of CNNs are criticized as ``black boxes" due to the difficulties in relating their outputs intuitively and quantitatively to the cosmological fields under investigation. To overcome this challenge, we present the \textit{Cosmological Correlator Convolutional Neural Network} (C3NN) --- a fusion of CNN architecture with
the framework of cosmological N-point correlation functions (NPCFs). We demonstrate that the output of this model can be expressed explicitly in terms of the analytically tractable NPCFs. Together with other auxiliary algorithms, we are able to open the ``black box" by quantitatively ranking different orders of the interpretable convolution outputs based on their contribution to classification tasks. As a proof of concept, we demonstrate this by applying our framework to a series of binary classification tasks using Gaussian and Log-normal random fields and relating its outputs to the analytical NPCFs describing the two fields. Furthermore, we exhibit the model's ability to distinguish different dark energy scenarios ($w_0=-0.95$ and $-1.05$) using N-body simulated weak lensing convergence maps and discuss the physical implications coming from their interpretability. With these tests, we show that C3NN combines advanced aspects of machine learning architectures with the framework of cosmological NPCFs, thereby making it an exciting tool with the potential to extract physical insights in a robust and explainable way from observational data.
\end{abstract}


\section{Introduction} 
\label{sec:intro}
\noindent In recent years, numerous machine learning methods have found applications in cosmology and astrophysics ranging from classification and regression tasks to acceleration of computational methods (see \citet{ml_and_cosmology_2022} for a recent review). Among the various machine learning techniques, Convolutional Neural Networks (CNNs) \citep{deeplearning_yann_lecun_2015} have been used extensively. Briefly, CNNs can compress a large dataset (e.g.~images) into several feature representations through a series of alternative linear and nonlinear transformations. The large number of parameters in a CNN model is typically determined by training the model to numerous simulated data that aim at reproducing actual scientific phenomena. In the context of astronomy and cosmology, these compressed features can be used for classification, such as searching for strong gravitational lensing systems \citep{des_cnn_search_strong_lensing_2022} and classifying different galaxy morphologies \citep{galaxy_morphology_cnn_2022}, or for inference analyses such as constraining parameters in various cosmological models \citep{Fluri_2019, fluri_kids1000, Lu_HSCY1_2023}, to name a few. 

However, this impressive development of CNNs have raised an important question: \textit{How to interpret the output feature representations of a CNN?} Ideally, we would like to establish a complete and controllable process in the application of CNN models in cosmology. However, most of the output features of a conventional CNN are notoriously difficult to interpret and hence they have often been termed as ``black boxes". In order to open such a ``black box", one good approach would be to connect the output feature representations from a CNN to either the analyzable statistical properties underlying the training data or visually understandable images that resemble the input training maps. If this can be achieved, we can understand what physical information in the training data the CNN pays most attention to in order to complete an assigned task. Moreover, the mechanism which the model adopts to extract this physical information can potentially provide us with certain knowledge that cannot be acquired by conventional CNN methods. 

To address this problem of interpretability in CNN models, different methods and architectures have been developed such as saliency maps \citep{saliency_maps_2013} (see e.g.~\citet{saliency_weak_lensing_2020} and \citet{saliency_21cm_2021} for a few applications), the three-dimensional CNN framework in \citet{3dcnn_luisa_cosmology}, variational autoencoders applied to find compressed representation of dynamical dark energy (DE-VAE) models \citep{DE-VAE} and so on. However, these techniques are either more visual-based checks which can only lead to qualitative conclusions or involve complex nonlinear transformations such that one cannot establish a straightforward mathematical relation between the output features and the input data. This can to a certain degree limit our interpretation of the physical information captured by CNNs.

In order to overcome this challenge, we introduce \textit{Cosmological Correlator Convolutional Neural Network} (C3NN) for cosmological analyses, adapted from the model initially proposed by \citet{Miles_ccnn_2021} in studies of correlated quantum matter. The output of this model can be explicitly related via a one-to-one correspondence to different orders of cosmological NPCFs. The whole framework of NPCFs and their applications in cosmology \citep{Peebles_1980, Bernardeau_2002} have been subject to extensive theoretical as well as practical developments over the last several decades. The 2-point correlation function (2PCF), or its Fourier space counterpart the power spectrum, is currently the most commonly used statistical method in many fields of cosmology such as weak gravitational lensing \citep{kids1000_cosmology_shear, des_y3_cosmic_shear_secco, hsc_y3_cosmic_shear}, galaxy clustering \citep{boss_galaxy_pk_cosmology, boss_galaxy_2pcf} and Cosmic Microwave Background (CMB) \citep{planck_2018_cosmology, planck_pr4_cosmology} analyses. Within the correlation function framework, active research has been done to push the theoretical modelling beyond 2PCF to access non-Gaussian cosmological information e.g.~in the studies of weak lensing and projected galaxy density fields such as \cite{3pcf_in_cosmology_takada_2003, 3pcf_in_weak_lensing_2003,Semboloni_2010,Friedrich_2018,i3pcf_2021_halder,3_times_i3pcf_galaxies,i3pcf_response,Gatti_2020,i3pcf_gong_2023,kids1000_2_3_shear_apertures,Heydenreich_2023, dhayaa_cgf_2023, Barthelemy_2024} to name a few. However, due to the difficulties in modelling and measurement and the treatment of systematic effects, extensive analyses of higher-order correlation functions in actual data remain challenging. Therefore, it is difficult to quantify the amount of additional cosmological information contained in 3-point, 4-point or even higher-order correlation functions. From this perspective, C3NN can provide some novel insights as we design it such that:
\begin{itemize}
    \item It can efficiently extract a series of statistical features which we call \textit{moments}. These moments are directly related to compressed NPCFs of the input data up to a desired order $N$ (user specified) by construction. C3NN relies on simulations for its training wherein one can include all the desired physics and systematic effects at the field level. Therefore, in extracting these output moments, C3NN side-steps the need for involved theoretical modelling of higher-order correlation functions. 
    \item At the same time, if one has a theoretical model for a given NPCF it can be directly used to interpret the corresponding order C3NN moment through straightforward mathematical relations as explained later.
    \item Using a certain post-processing numerical method called regularization path analysis, we can easily understand the relative importance of the output moments in tasks such as binary classification, e.g.~to distinguish between different cosmological model scenarios beyond $\Lambda\rm CDM$. This in turn can provide guidance to perform the corresponding NPCF analyses in real observations.
    \item Finally, the filter weights that are learned by C3NN can be used to pick out the specific configurations within a given NPCF. This is extremely desirable as it also allows us to understand not only the relative importance of a given order moment but also the configurations which carry the most distinct information for performing an assigned task.
\end{itemize}
As this is the first time we introduce C3NN to cosmology, we only consider the task of binary classification with two-dimensional simulated maps in this work. However, the same architecture can be exploited in cosmological parameter inference or classification with three-dimensional simulated boxes, etc. We leave such topics to future works. This paper is organized as follows: In Section \ref{sec:model_architecture}, we describe the architecture of C3NN. We then give clear mathematical expressions for the output moments and explicitly show their relations to the conventional correlation functions in cosmology. We also discuss the differences among C3NN, CNN and scattering transform. In the following sections \ref{sec:gaussian_fields} and \ref{sec:gaussian_lognormal_fields}, we present as a proof of concept, two test cases of C3NN classifying between (i) two different Gaussian random fields, and (ii) between Gaussian and Log-normal random fields. Then in Section \ref{sec:n_body_results} we further apply C3NN to distinguish between different dark energy scenarios using N-body simulated weak lensing convergence maps and discuss the physical implications coming from its interpretability. Finally, we summarise and conclude in Section \ref{sec:summary_and_conclusion}.

\section{C3NN model architecture and interpretability} 
\label{sec:model_architecture}
\begin{figure}[bt!]
\includegraphics[width=\textwidth]{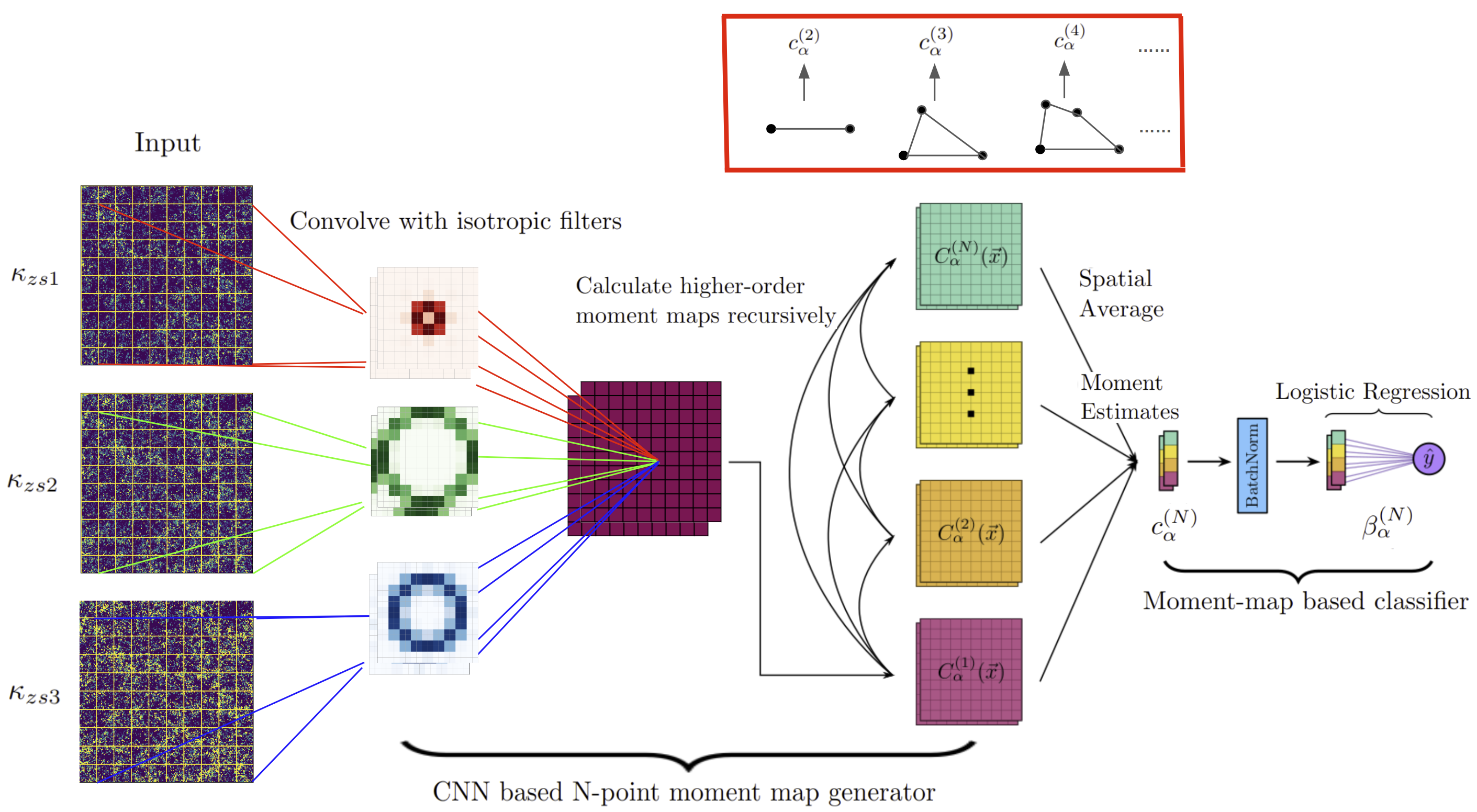}
\caption{A schematic illustration of the C3NN architecture. The input data can be multi-channel 2D maps (e.g.~weak lensing convergence maps with multiple distinct source redshift bins). The dimensionality of the input data tensor is $(D, C, W, H)$ where $D$ is the number of simulation realizations, $C$ the number of channels and $W$ and $H$ describing the width and height (in pixels) of each individual map. This input data tensor is then convolved with isotropic filters built with \texttt{ESCNN} (see details in Section \ref{sec:model_architecture}) in order to produce the first moment map $C_{\alpha}^{(1)}$. With this single convolution operation, one can calculate higher-order moment maps of the input field (up to order $N$ which is user defined) recursively (see Section \ref{sec:model_architecture} for details). The maps are then spatially averaged in order to obtain compressed scalar quantities that we call moments $c_{\alpha}^{(N)}$. As shown in the top red rectangular box, these $c_{\alpha}^{(N)}$ are associated with the corresponding orders of NPCFs. The calculated moments can then be passed through the classifier part of the model to perform a binary classification task. Part of this figure has been adapted from \citet{Miles_ccnn_2021}.} 
\label{fig:ccnn_architecture}
\end{figure}

\noindent In this section, we present the architecture of C3NN whose overall structure is shown in Figure \ref{fig:ccnn_architecture}. We split it into two parts and name them as \textit{CNN based N-point moment map generator} and \textit{moment-map based classifier}, respectively.\\
\\
\noindent \textbf{CNN based N-point moment map generator:} The generator part of the architecture performs convolution operations recursively in order to produce moment maps $C_{\alpha}^{(N)}$ (of order $N$) which we define as follows:
\begin{equation}
    \label{eq:C_N_def}
    \begin{split}
        C_{\alpha}^{(1)}(\textbf{x}) &= \sum\limits_{\textbf{a},k}w_{\alpha,k}(\textbf{a})S_{k}(\textbf{x}+\textbf{a}) \\ 
        C_{\alpha}^{(2)}(\textbf{x}) &= \frac{1}{2!}\left[\sum\limits_{(\textbf{a},k) \neq (\textbf{a}_1,k^{\prime})}w_{\alpha,k}(\textbf{a})w_{\alpha,k^{\prime}}(\textbf{a}_1)S_{k}(\textbf{x}+\textbf{a})S_{k^{\prime}}(\textbf{x}+\textbf{a}_1)\right] \\
        \dots \\
        C_{\alpha}^{(N)}(\textbf{x}) &= \frac{1}{N!}\left[\sum\limits_{(\textbf{a},k) \neq\dots\neq (\textbf{a}_N,k_{N})}\prod\limits_{j=1}^{N}w_{\alpha,k_{j}}(\textbf{a}_j)S_{k_{j}}(\textbf{x}+\textbf{a}_j)\right] \ .
    \end{split}
\end{equation}
Here $w$ is the filter weight and $S$ denotes the input map (e.g.~a pixelized image of a 2D field). During training, we enforce $w$ to take its absolute value to aid interpretation. Position vectors $\textbf{a}$, $\textbf{a}_1$ $\dots$ $\textbf{a}_N$ of the filter weight run over every pixel. The factorial factor removes repeated counting of the same combination of filter weights and input map pixels. Index $\alpha$ represents different filters while index $k$ goes through different channels. Each filter can have several channels which correspond to those in the input data, e.g.~different tomographic bins of an observable. From the above equations, we see clearly that for $N \geq 2$, $C_{\alpha}^{(N)}(\textbf{x})$ is the weighted summation of all possible pixel configurations within the filter at position $\mathbf{x}$. For $N=1$, $C_{\alpha}^{(1)}$ represents the usual convolution of the input data with the CNN filter. In the summation, we strictly exclude zero-lag contributions at any order. The summation over $k$ enables C3NN to measure cross-correlations among different channels of the input data. It is exactly relying on this property that we can straightforwardly implement C3NN within the context of a tomographic analysis as we will discuss later in section.~\ref{sec:n_body_results}. Moreover, in the case when one provides different observables as input to the different channels, C3NN would capture cross-correlations among these quantities, e.g. the 3$\times$2-point probes of cosmic shear and galaxy clustering that are routinely analysed in galaxy imaging surveys \citep{kids1000_3_cross_2pt_2021, des_y3_3_cross_2pt_2022}. We defer such investigations to future works.

One difficulty to compute moment maps in Eq.~\ref{eq:C_N_def} is the increasing computational cost along with the order $N$. The cost can be approximated as $\mathcal{O}((KP)^N)$ per input map pixel where $K$ is the number of input data channels and $P$ is the number of pixels in the convolutional filter. To solve this problem, \citet{Miles_ccnn_2021} (see their section.~S II) have proved that $C_{\alpha}^{(N)}$ can be calculated using a recursive formula:
\begin{equation}
    \label{eq:C_N_recursive}
    C_{\alpha}^{(N)}(\textbf{x}) = \frac{1}{N}\sum\limits_{\ell=1}^{N}(-1)^{\ell-1}\left(\sum\limits_{\textbf{a},k}w_{\alpha,k}^{\ell}(\textbf{a})S_{k}^{\ell}(\textbf{x}+\textbf{a})\right)C_{\alpha}^{(N-\ell)}(\textbf{x}) \ ,
\end{equation}
where the power operation is taken on every pixel and we set $C_{\alpha}^{(0)}(\textbf{x})$ to be 0. Through this relation, the computational cost decreases to $\mathcal{O}(N^2KP)$. Instead of direct computation at each order of the moment map, we can now convolve the input only once to produce $C_{\alpha}^{(1)}$. Then $C_{\alpha}^{(2)}$ can be calculated from $C_{\alpha}^{(1)}$. Similarly $C_{\alpha}^{(3)}$ can be calculated from $C_{\alpha}^{(1)}$ and $C_{\alpha}^{(2)}$ and so on via Eq.~\ref{eq:C_N_recursive}. This recursive relation holds until we truncate at an order $N$.\\
\\
\noindent \textbf{Moment-map based classifier:} This classifier part of the architecture starts with compressing the $C_{\alpha}^{(N)}$ maps obtained from the generator into scalars $c_{\alpha}^{(N)}$ by taking the spatial average over the maps:
\begin{equation}
    \label{eq:c_n_def}
    c_{\alpha}^{(N)} = \frac{1}{N_{\rm pix}}\sum\limits_{\textbf{x}}C_{\alpha}^{(N)}(\textbf{x}) \ ,
\end{equation}
where $N_{\rm pix}$ is the total number of pixels in the input map $S$. Based on Eqs.~\ref{eq:C_N_def} and \ref{eq:c_n_def}, we can explicitly show the relation between $c_{\alpha}^{(N)}$ and the corresponding NPCF both at the discrete filter pixel level and in the continuous limit. As an example, when $N=2$, we have:
\begin{equation}
    \label{eq:c2_and_2pcf}
    \begin{split}
        c_{\alpha}^{(2)} &= \frac{1}{N_{\rm pix}}\sum\limits_{\textbf{x}}\frac{1}{2!}\left[\sum\limits_{(\textbf{a},k) \neq (\textbf{a}_1,k^{\prime})}w_{\alpha,k}(\textbf{a})w_{\alpha,k^{\prime}}(\textbf{a}_1)S_{k}(\textbf{x}+\textbf{a})S_{k^{\prime}}(\textbf{x}+\textbf{a}_1)\right] \\
        &= \frac{1}{2!}\sum\limits_{(\textbf{a},k) \neq (\textbf{a}_1,k^{\prime})}w_{\alpha,k}(\textbf{a})w_{\alpha,k^{\prime}}(\textbf{a}_1)\left[\frac{1}{N_{\rm pix}}\sum\limits_{\textbf{x}}S_{k}(\textbf{x}+\textbf{a})S_{k^{\prime}}(\textbf{x}+\textbf{a}_1)\right] \\
        &= \frac{1}{2!}\sum\limits_{(\textbf{a},k) \neq (\textbf{a}_1,k^{\prime})}w_{\alpha,k}(\textbf{a})w_{\alpha,k^{\prime}}(\textbf{a}_1)\hat{\xi}_{kk^{\prime}}(\textbf{a}_1-\textbf{a}) \\
        &= \frac{1}{2!}\sum\limits_{(\textbf{a},k) \neq (\textbf{a}+\textbf{r},k^{\prime})}w_{\alpha,k}(\textbf{a})w_{\alpha,k^{\prime}}(\textbf{a}+\textbf{r})\hat{\xi}_{kk^{\prime}}(\textbf{r}) \ ,
    \end{split}
\end{equation}
where $\hat{\xi}_{kk^{\prime}}$ is the volume average estimator for auto/cross 2PCFs at a separation $\textbf{r}$ between channels $k$ and $k'$. Note that both $\textbf{a}$ and $\textbf{a}+\textbf{r}$ in the last line should be strictly contained within the filter. Therefore, the scale of correlation functions that C3NN can probe is restricted by the filter size. Similarly, $c_{\alpha}^{(3)}$ can be written as:
\begin{equation}
    \label{eq:c3_and_3pcf}
    c_{\alpha}^{(3)} = \frac{1}{3!}\left[\sum\limits_{(\textbf{a},k) \neq (\textbf{a}+\textbf{r},k_1) \neq (\textbf{a}+\textbf{r}^{\prime},k_2)}w_{\alpha,k}(\textbf{a})w_{\alpha,k_1}(\textbf{a}+\textbf{r})w_{\alpha,k_2}(\textbf{a}+\textbf{r}^{\prime})\hat{\zeta}_{kk_1k_2}(\textbf{r},\textbf{r}^{\prime},-\textbf{r}-\textbf{r}^{\prime})\right] \ ,
\end{equation}
where $\hat{\zeta}_{kk_1k_2}$ is the volume average estimator for auto/cross 3-point correlation functions (3PCFs) where closed triangles can be formed by 3 pixels within the filter. To complete the demonstration, we consider the calculation of $c_{\alpha}^{(2)}$ in the limit of a continuous filter function:
\begin{equation}
    \label{eq:c2_continuous_limit}
        c_{\alpha}^{(2)} = \frac{1}{2!}\lim_{\textbf{q}\to 0} \sum\limits_{k,k^{\prime}}\int {\rm d}^2\textbf{a} \ W_{\alpha,k}(\textbf{a})\int_{\textbf{r} \neq \textbf{0}} {\rm d}^2\textbf{r} \ W_{\alpha,k^{\prime}}(\textbf{a}+\textbf{r})\hat{\xi}_{kk^{\prime}}(\textbf{r}) \ ,
\end{equation}
where $\textbf{q}$ is the separation between nearest neighboring pixels in the filter and $W_{\alpha,k}$ here acts as a continuous two-dimensional kernel function that has a sharp boundary outside which its values are always 0. This is to take the finite filter size into consideration. A similar calculation for $c_{\alpha}^{(3)}$ follows analogously. These explicitly show the relation between $c_{\alpha}^{(N)}$ and NPCFs thereby robustly identifying them as interpretable summary statistics. 

We note that $c_{\alpha}^{(N)}$ is an integration over all the configurations of a given order NPCF and hence we call it a \textit{N-point moment}. It is interesting to compare $c_{\alpha}^{(n)}$ to the mass aperture moments \citep{3rd_order_mass_aperture_schneider_2005, 2nd_3rd_mass_aperture_kilbinger_2005, sven_mass_aperture_2023}. Their expressions are very similar. The difference is that mass aperture moments exploit analytical kernel functions such as compensated filters \citep{compensation_filter_crittenden_2002} which can have different scale radii while ours are optimized by the training process and the same filter is applied to all input data pixels. One point to note is that it is difficult to write down expressions for mass aperture moments beyond 3rd order as it requires accurate modelling for trispectrum and so on. On the other hand, C3NN can measure a whole sequence of $c_{\alpha}^{(n)}$ up to an arbitrary order. Another perspective to understand $c_{\alpha}^{(n)}$ is that since it explicitly excludes the measurement of self-correlation, it is also different from the conventional definition of $N$th-order moment which is composed of zero-lag correlation functions. For simplicity, in the following texts, we will mention $c_{\alpha}^{(n)}$ as moments unless otherwise specified but the real definition should be made clear.

After building all the $c_{\alpha}^{(N)}$ up to a given order $N$, we group them together to form a feature vector $\textbf{c} = \{c_{\alpha}^{(1)}, c_{\alpha}^{(2)}, c_{\alpha}^{(3)},..., c_{\alpha}^{(N)}\}$. Then the vector goes through a batch normalization before the model passes it through a logistic classifier for a final probabilistic prediction $\hat{y}$ to differentiate two classes:
\begin{equation}
    \label{eq:logistic_prediction}
    \hat{y} = \frac{1}{1+{\rm e}^{-\boldsymbol{\beta}\cdot\boldsymbol{c}+\epsilon}} \ ,
\end{equation}
where $\boldsymbol{\beta}$ is a trainable vector composed of coefficients $\beta_{\alpha}^{(N)}$ which have a one-to-one correspondence to $c_{\alpha}^{(N)}$. The other trainable parameter is the bias scalar $\epsilon$. When $-\boldsymbol{\beta}\cdot\boldsymbol{c}+\epsilon=0$, the probabilistic prediction would always be $0.5$ which implies a random classification between two classes. From that naturally the hyperplane $-\boldsymbol{\beta}\cdot\boldsymbol{c}+\epsilon=0$ acts as the decision boundary in the high $N$-dimensional space. C3NN would classify the feature vector $\textbf{c}$ that falls to either side of the decision boundary into a specific class. In other words, if $\hat{y} \in [0,0.5)$, the model would predict the input data to belong to one class. Otherwise, the input data would be considered as the other class. 
\begin{figure}[bt!]
\centering
\includegraphics[scale=0.18]{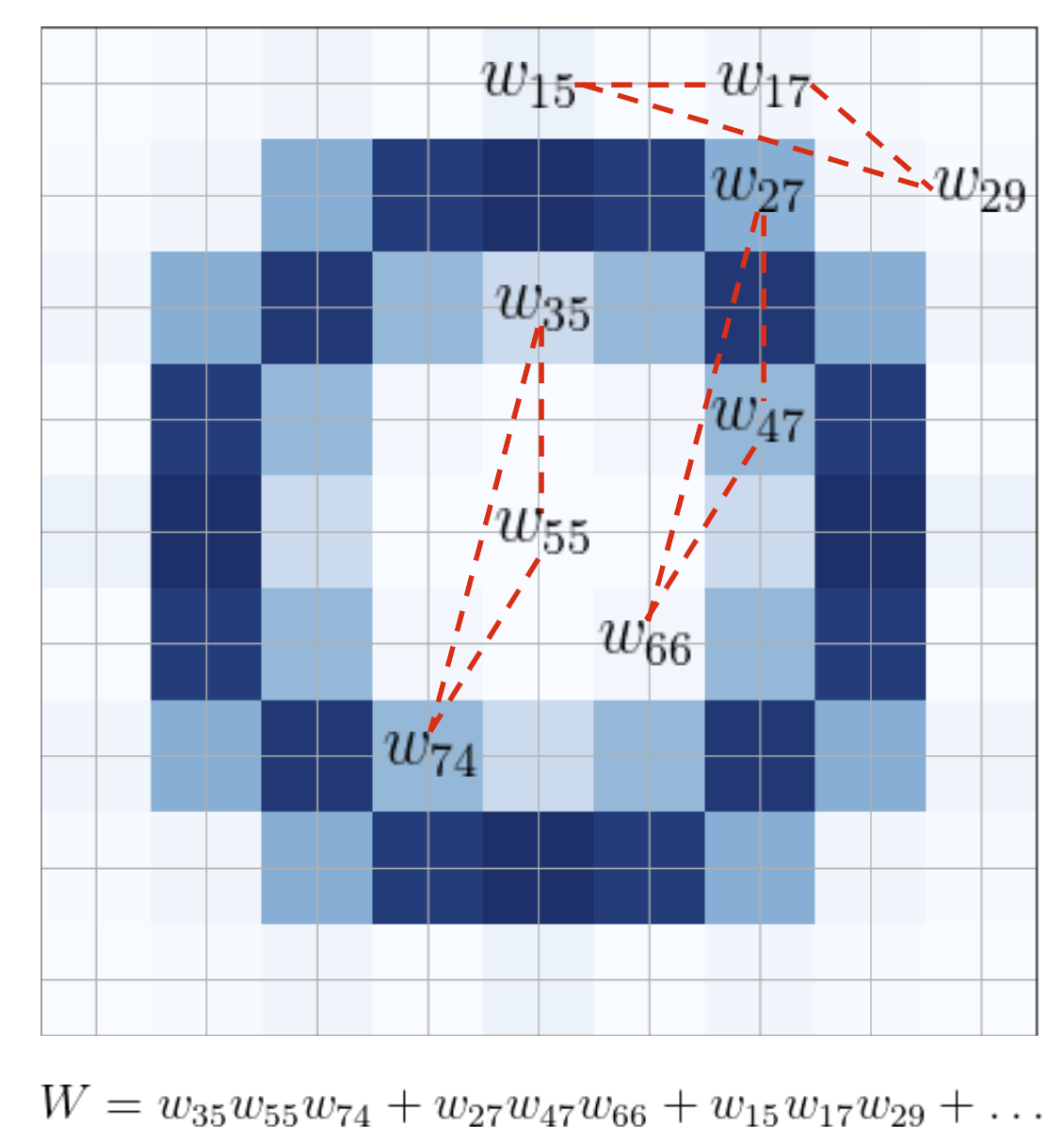}
\caption{A demonstration of how C3NN filter weights can be analyzed: Suppose we have $\beta_{1}^{(3)}$ first activated in the regularization path analysis, we can then check the filter and quantify the relative importance of different configurations of 3PCF that $c_{1}^{(3)}$ is composed of. Here we assume that the trained rotationally symmetric filter is of size $9 \times 9$ pixels and it only has one channel. Each intersection between a vertical and a horizontal line represents the center of a pixel. The trained weight is denoted by $w_{\rm ij}$ and any distinct triplet would form a closed triangle that is included in the measurement of $c_{1}^{(3)}$. As an example, the three triangles drawn in red dashed lines are weighted as $w_{35}$$w_{55}$$w_{74}$, $w_{27}$$w_{47}$$w_{66}$ and $w_{15}$$w_{17}$$w_{29}$ respectively according to Eq.~\ref{eq:C_N_def}. Since they are equivalent triangles, their weights can be summed, along with all the other equivalent triangles that can be found within the filter. The summation is the total weight $W$ of this particular triangle configuration. We want to emphasize that all configurations of the 3PCF mentioned here are in real space, which are different from triangles in Fourier space often discussed in the context of bispectrum studies in literature. The same approach can be applied to 2-point separations, quadrilaterals, etc.} 
\label{fig:filter_analysis_demonstration}
\end{figure}

To further develop the interpretability, we adopt the a two-round training strategy: In the first round, we train the whole C3NN exhibited in Figure \ref{fig:ccnn_architecture}. In this case, we use the binary cross-entropy as the loss function with an extra L1 regularization term on the filter weights:
\begin{equation}
    \label{eq:1st_training_L1_loss}
    L_{\rm 1st}(y,\hat{y}) = -y{\rm log}\hat{y} - (1-y){\rm log}(1-\hat{y}) + \gamma\sum\limits_{\alpha,k, \textbf{a}}w_{\alpha,k}(\textbf{a}) \ ,
\end{equation}
where $\gamma$ is the regularization strength. This regularization term helps to turn off unnecessary pixels \citep{lattice_ft_explanable_ml_2020} which aid our later interpretation of the filter weights. The resulting C3NN model can be directly used to perform binary classification tasks. In the second round, we freeze the filter weights from the previous round and use the moment map generator to produce moment maps. We then use those maps to re-train $\beta_{\alpha}^{(N)}$ coefficients. It is important to note that the value of $\beta_{\alpha}^{(N)}$ indicates the relative importance of $c_{\alpha}^{(N)}$ in its contribution to the classification, e.g. from Eq.~\ref{eq:logistic_prediction} when $\beta_{\alpha}^{(N)}=0$, its corresponding $c_{\alpha}^{(N)}$ would be irrelevant in C3NN's decision making. Based on this property, we use a different L1 regularization term in the binary cross-entropy loss during the second round of training:
\begin{equation}
    \label{eq:2nd_training_L1_loss}
    L_{\rm 2nd}(y, \hat{y}) = -y{\rm log}\hat{y} - (1-y){\rm log}(1-\hat{y}) + \lambda\sum\limits_{\alpha,n}|\beta^{(n)}_{\alpha}| \ ,
\end{equation}
where $\lambda$ is the regularization strength. We train the $\beta_{\alpha}^{(N)}$ coefficients iteratively with gradually decreasing $\lambda$ values. This is a process of feature selection using the so called regularization path analysis \citep{least_angle_regression_2004, tang2014feature}. Initially, when $\lambda$ is large, all $\beta_{\alpha}^{(N)}$ values are close to 0 in order to minimize $L_{\rm 2nd}$ and therefore the model does not possess any classification power. After $\lambda$ decreases to a certain value, the most important feature represented by a given $\beta_{\alpha}^{(N)}$ would be activated since at this stage the increase in the loss from $\lambda\sum\limits_{\alpha,n}|\beta^{(n)}_{\alpha}|$ would be over-compensated by the decrease from $-y{\rm log}\hat{y} - (1-y){\rm log}(1-\hat{y})$ and thus minimize the total loss. As $\beta_{\alpha}^{(N)}$ is coupled to $c_{\alpha}^{(N)}$, the sequence of the activation of $\beta_{\alpha}^{(N)}$ can provide us insights into the relative importance of different moments.

Once we find the sequence of the $\beta_{\alpha}^{(N)}$ activation, we can check the filter weights that are trained during the first round. For each corresponding moment $c_{\alpha}^{(N)}$ measured by a specific filter, we can quantify the relative importance of a particular NPCF configuration. The method is demonstrated in Figure \ref{fig:filter_analysis_demonstration}. As an aside, we note here that the smallest configuration of the correlation function at a given order is determined by the pixel resolution of the input map. This method implies that we can not only infer a rank of moments based on their relative importance in classification, but also a rank of correlation function configurations based on the relative weights within each moment.

The filters used in the convolution are constructed in such a way that its weights after training are rotationally symmetric with respect to the central filter pixel, i.e.~weights that have the same distance to the central pixel are identical. We impose this symmetry because the spatially averaged moments $c_{\alpha}^{(N)}$ should be invariant with respect to rotations of the training maps. From the perspective of the cosmological principle, this ensures that our model obeys the isotropy of the Universe. In practice, we implement this in our filters using the equivariant steerable convolutional neural network\footnote{currently hosted at: \url{https://github.com/QUVA-Lab/escnn}}(\texttt{ESCNN}) \citep{e2cnn, escnn_cesa_2022} architecture. The requirement of rotational invariance imposed on the filter naturally decreases the number of free weights to be trained and therefore makes the training procedure more numerically efficient as it needs fewer training data.

Besides the actual trainable parameters such as filter weights and $\beta_{\alpha}^{(N)}$, we also consider the hyper-parameters within the model. Parameters like the learning rate or regularization strength cannot update themselves during the training but their values can affect the model performance. Therefore, in practice we use \verb|Optuna|\footnote{currently hosted at \url{https://optuna.org/}} \citep{optuna_2019} as the framework for optimizing these hyper-parameters.

It is necessary to point out, based on what we have already discussed, the differences between C3NN and some other AI methods which from a first glance can look similar to our model. For example, in conventional CNNs used in cosmological research \citep{Fluri_2019, fluri_kids1000, Lu_HSCY1_2023}, there are several convolutional layers, each of which is followed by a nonlinear activation function. This type of information extraction compresses the input data into some compact feature vectors which are then used by the model to undertake tasks such as classification or regression. Although very powerful, one disadvantage of this conventional CNN framework is that it is generally hard to interpret the reasons why the CNN compresses the input data in the way it does in order to train its model parameters. On the other hand, our C3NN architecture has only one convolution layer. We discard the general nonlinear transformations and instead replace them with the expansion of moment-maps via a recursive relation. In other words, our C3NN architecture is simpler than the conventional multi-layered CNNs. Meanwhile, the final compressed feature vectors from C3NN can be mathematically related to the concept of NPCFs as discussed above, which are summary statistics with great theoretical tractability and are the basis of statistical analyses in cosmology over the past few decades. 

Scattering transforms \citep{Cheng_st_2020, Cheng_st_2021} are also different from C3NN in the sense that it uses pre-determined wavelet filters. C3NN filters on the other hand are optimized during the training process. Moreover, unlike C3NN, the final compressed coefficients of the scattering transforms (no matter up to which order they are computed at) do not represent the corresponding NPCF that is contained in the C3NN output $c_{\alpha}^{(N)}$.

\section{Proof of concept tests}
\label{sec:proof_of_concept}
In this section, we test C3NN on two binary classification tasks: (i) Gaussian random fields with different correlation lengths (Section \ref{sec:gaussian_fields}), and, (ii) Gaussian and Log-normal random fields with the same power spectrum but different higher-order spectra (Section \ref{sec:gaussian_lognormal_fields}). Using these two tests we also demonstrate how the regularization path analysis and filter weights analysis relates to our model's interpretability as mentioned in the previous section.

\subsection{Test case of Gaussian random fields}
\label{sec:gaussian_fields}
\begin{figure}[bt!]
\centering
\includegraphics[scale=0.6]{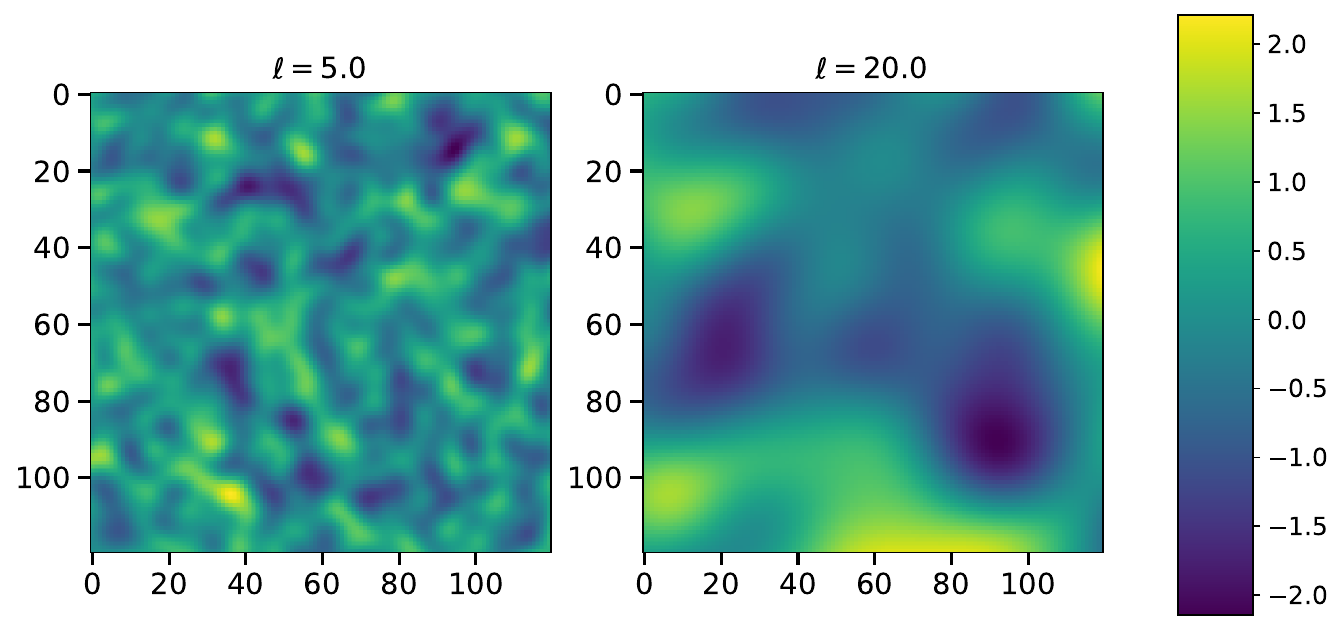}
\caption{Two classes of Gaussian random fields generated by \texttt{GSTools} where we show one example for each. Both classes have the same variance amplitude of 1.0 and a map size of $120 \times 120$ pixels. They differ from each other on the correlation length, with one having $\ell = 5.0$ (left panel) and the other having $\ell = 20.0$ (right panel), with the correlation lengths specified in unit of pixels. The color bar to the right shows the scale of the field amplitude.} 
\label{fig:grf_gstools}
\end{figure}

As the first proof of concept, we apply C3NN to distinguish two classes of Gaussian random fields, each with a different correlation lengths. We use the \texttt{GSTools}\footnote{https://geostat-framework.readthedocs.io/projects/gstools/en/stable/} \citep{gstools_2022} framework to generate these Gaussian random fields as square 2D maps. While keeping the variance amplitude of the two fields the same, we vary their field correlation lengths $\ell$ as shown in Figure \ref{fig:grf_gstools}
where the realization in the left panel (smaller correlation length) clearly shows more fluctuations on small scales than the right one (larger correlation length). 

We prepare 5000 maps for each class, and every map only has 1 channel. We split this sample into 4500 for training and 500 for validation. Some model parameters including filter number, filter size and the highest correlation order to measure are predetermined to be 1, $31 \times 31$ pixels and 3, respectively. The filter size is chosen to be large enough to capture the correlation scales of both training classes. We then use \verb|Optuna| for a grid search to optimize the hyper-parameters in the model such that their combination can maximize the validation accuracy. The results are shown in Table \ref{tab:ccnn_grf}.
\begin{table}[ht!]
    \centering
    \begin{tabular}{|c|c|c|c|c|}
    \hline
      parameter & $\gamma$ & learning rate (lr) & learning rate decaying ratio ($\phi$) & optimizer \\
    \hline
      value & 2.33 & 0.47 & 0.02 & ``Adam" \\
    \hline
    \end{tabular}
    \caption{Four optimized hyper-parameters in C3NN trained on two classes of Gaussian random fields with different correlation lengths (see Figure \ref{fig:grf_gstools}). $\gamma$ is the regularization strength in Eq.~\ref{eq:1st_training_L1_loss}. We use a learning rate scheduler which decays the initial learning rate (lr) of each parameter group by a factor $\phi$ after every epoch. We simultaneously also search for the type of optimizer which performs the best and we obtain Adam, which is derived from adaptive moment estimation \citep{adam_2014}.}
    \label{tab:ccnn_grf}
\end{table}
The model with the above hyper-parameter values can return a validation accuracy of $99.2\%$ after 100 training epochs.

\begin{figure}[bt!]
\centering
\includegraphics[width=0.4\linewidth]{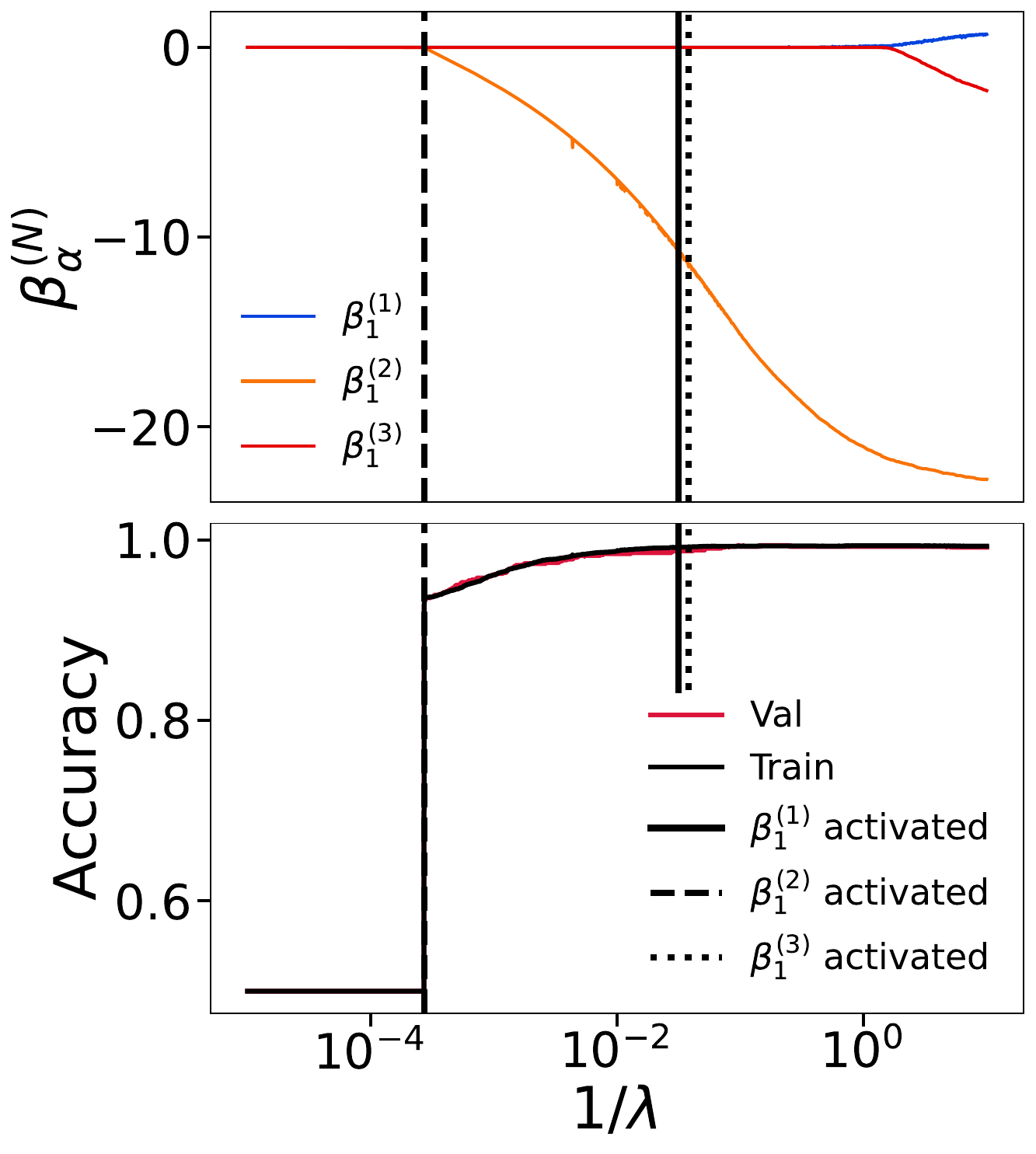}
\includegraphics[width=0.5\linewidth]{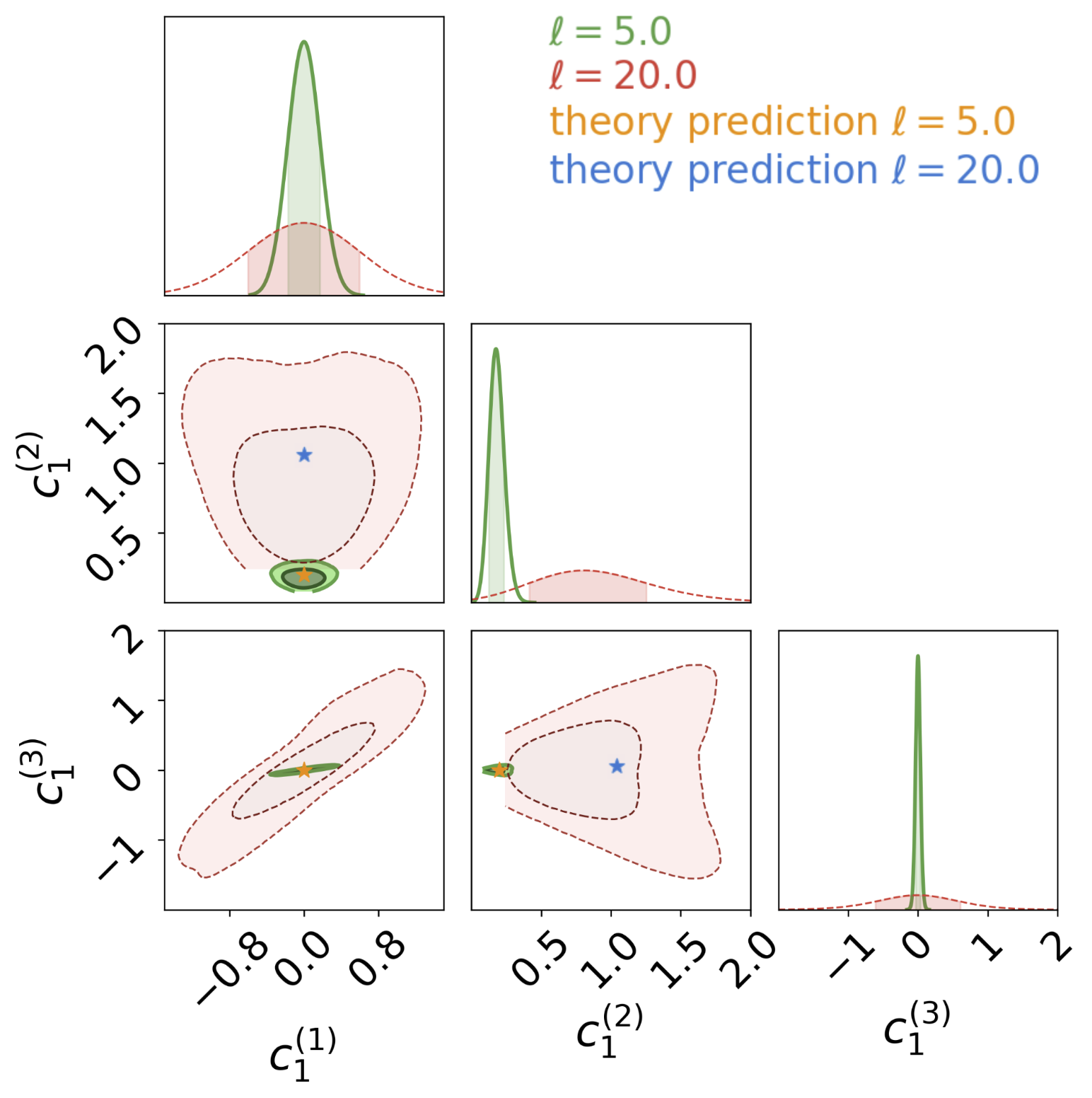}
\caption{\textit{Left}: The regularization path analysis result of C3NN trained on two classes of Gaussian random fields with correlation lengths $\ell = 5.0$ and $20.0$ (see Figure \ref{fig:grf_gstools}). The upper panel shows the evolution of the $\beta_{\alpha}^{(N)}$ coefficients along with the decrease in the regularization strength $\lambda$ in Eq.~\ref{eq:2nd_training_L1_loss} (an increase in $1/\lambda$). The bottom panel shows the corresponding changes of training and validation accuracy. Different types of vertical lines indicate the first $\lambda$ value at which each $\beta_{\alpha}^{(N)}$ becomes non-zero (i.e.~gets activated). \textit{Right}: Contours of moment $c_{\alpha}^{(N)}$ distributions mapped by passing the full training dataset into the moment map generator and spatial averaging after the first round of training. The diagonal subplots show the marginalized distributions of each order of moments from the two training classes. The green contours represent the moment distribution of the Gaussian random field with $\ell=5.0$ and the red contours represent that with $\ell=20.0$. The orange and blue stars are the theoretical predictions of different $c_{\alpha}^{(N)}$ for the two classes based on the trained filter weights and analytical expressions of correlation functions for a Gaussian random field.}
\label{fig:path_analysis_grf}
\end{figure}

We implement the regularization path analysis as discussed in the previous section and results are shown in the left panel of Figure \ref{fig:path_analysis_grf}. 
It is clear that the 2nd-order moment $c_{1}^{(2)}$ is the most important feature in this classification as $\beta_{1}^{(2)}$ is activated first. Once it is activated, both the training and validation accuracy increase from $50\%$, i.e. no classification power, to more than $90\%$. Afterwards, both the training and validation accuracy experience a slow increase while the value of $\beta_{1}^{(2)}$ goes through a fine tuning along the regularization path. At the activation of $\beta_{1}^{(1)}$ and $\beta_{1}^{(3)}$, the validation accuracy has almost reached $100\%$ already. Therefore we interpret these two activations as overfitting. This result is consistent with our expectation that the two classes of training data are completely distinguished based on the 2nd-order moment $c_{1}^{(2)}$ since for a Gaussian random field, one can characterize it completely by its expectation value and covariance. Since the two classes of training data have the same expectation value, the dominant power to distinguish them should come from the quantity which relates to the covariance i.e.~the 2nd-order moment $c_{1}^{(2)}$. The right panel in Figure \ref{fig:path_analysis_grf} shows that C3NN successfully captures this statistical property by using its trained filter to map the two classes of Gaussian random fields into two distinct distributions in the \{$c_{1}^{(n)}$\} space, particular along the marginalized dimension of $c_{1}^{(2)}$. It is based on these different distributions that the classifier learns the decision boundary between the two classes.

\begin{figure}[bt!]
    \centering
    \includegraphics[scale=0.35]{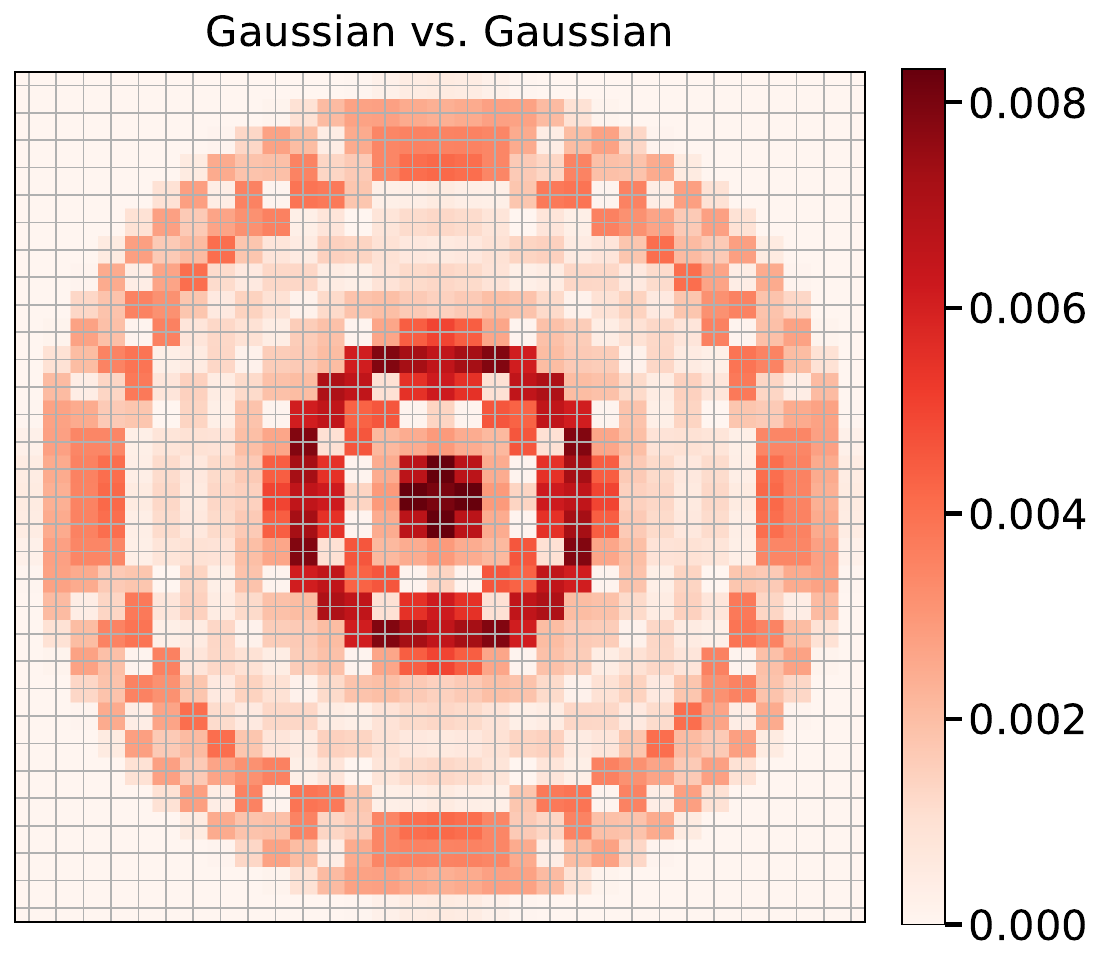}
    \includegraphics[scale=0.35]{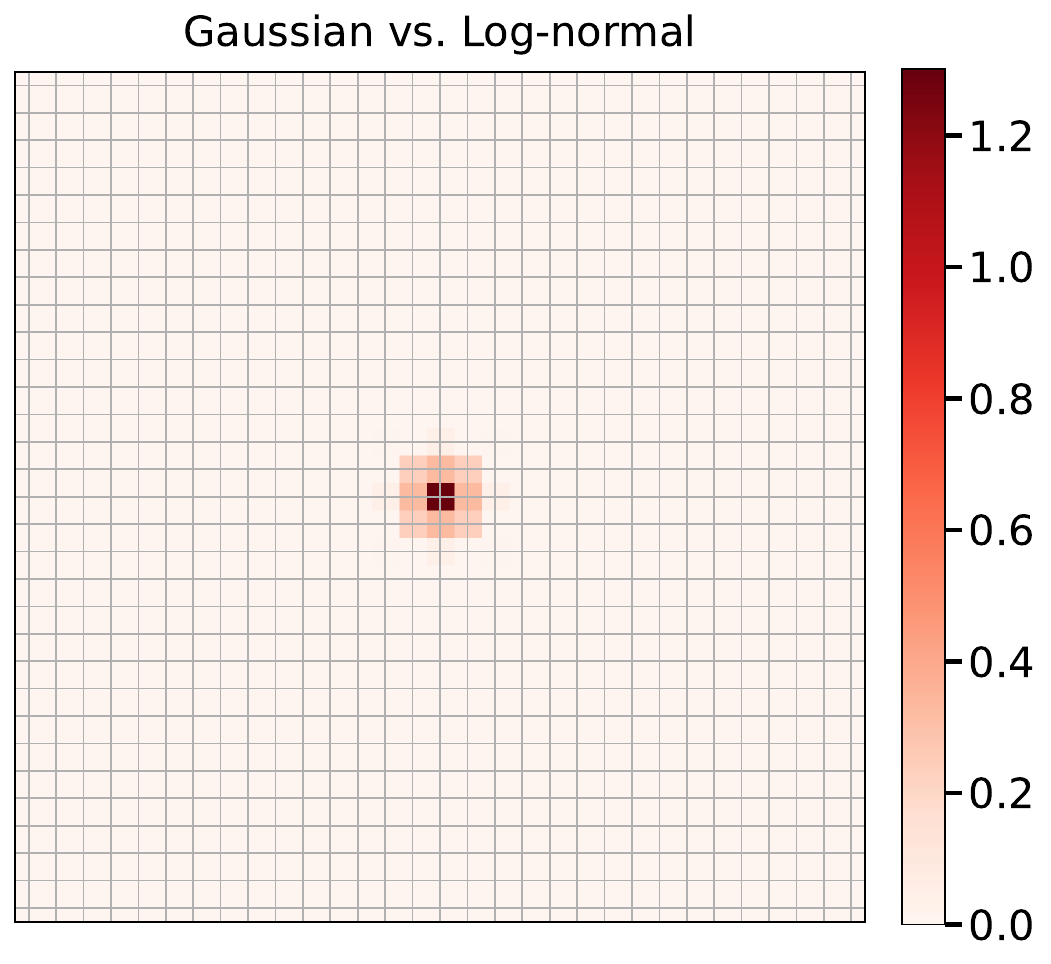}
    \caption{\textit{Left}: The trained filter of C3NN classifying two Gaussian random fields with different correlation length (see Figure \ref{fig:grf_gstools}). \textit{Right}: The trained filter of C3NN classifying Gaussian and Log-normal random fields with the same power spectrum but different higher order NPCFs (see Figure \ref{fig:trained_filter_grf_lognormal}). In both cases the filter is of size $31 \times 31$ pixels and all weights are non-negative. Both color bars show the scales of the filter weights.}
    \label{fig:trained_filter_grf_lognormal}
\end{figure}

Looking at the filter in the left panel of Figure \ref{fig:trained_filter_grf_lognormal}
we see that the learned filter weights show a rotationally symmetric pattern as required by our architecture. There are two prominent annuli on different scales. This indicates that there is a correspondence between the annulus scales in the filter and the characteristic scales in the Gaussian random fields. The outer annulus has a much larger radius than the correlation length $5.0$, hence it can measure correlation function signals that only appear in the training data class with $\ell=20.0$. The inner annulus, which has a radius of approximately 5 pixels, is more efficient in extracting correlation function signals from the training data class with $\ell=5.0$. As $\beta_{1}^{(2)}$ is first activated, we can quantify the relative importance of 2PCF with different separations following the approach described in Section \ref{sec:model_architecture} and Figure \ref{fig:filter_analysis_demonstration}. The most important 2PCF which contributes to the classification is exactly the one with a separation of 5.0 pixels with a corresponding total weight is 0.028. The following two most important 2PCFs are with separations 8.06 and 12.04. Their associated total weights are 0.024 and 0.023 respectively.

The above filter can even be used to compute $\{c_{1}^{(1)}, c_{1}^{(2)}, c_{1}^{(3)}\}$ analytically. We know theoretically for the training Gaussian random fields, their expectation value and all connected 3PCFs should be 0.0. Therefore $c_{1}^{(1)}$ and $c_{1}^{(3)}$ for both training data classes would be 0.0. For $c_{1}^{(2)}$, first we select all possible 2PCF configurations within the filter and calculate the total weight for each of them, then by following Eq.~\ref{eq:c2_and_2pcf} we combine these total weights with the corresponding 2PCF amplitudes that can be computed analytically for Gaussian random fields. The equation we use in this case is the one used by \texttt{GSTools} to create the Gaussian maps \citep{webster2007geostatistics}:
\begin{equation}
    \label{eq:2pcf_gstools}
    \gamma(r) = \sigma^2\left(1 - {\rm exp}\left[-\left(\frac{sr}{\ell}\right)^2\right] \right) \ ,
\end{equation}
where $\gamma$ represents the 2PCF amplitude at a given separation $r$. The standard rescale factor is $s=\sqrt{\pi}/2$. The variance amplitude is denoted by $\sigma^2$ and $\ell$ is the correlation length of the Gaussian random field. The results are $c_{1}^{(2)}=0.199$ for $\ell=5.0$ and $1.049$ for $\ell=20.0$. They are marked by orange and blue stars in the right panel of Figure \ref{fig:path_analysis_grf} and are well contained within the 1$\sigma$ confidence interval of the distributions mapped by C3NN from the input maps of the two classes. This confirms that the moments $c_{\alpha}^{(N)}$ measured by C3NN can be interpreted in terms of correlation functions.

\subsection{Test case of Gaussian and Log-normal distributed weak lensing convergence fields}
\label{sec:gaussian_lognormal_fields}

\noindent In cosmology the central field of interest is the 3D matter density contrast field $\delta^{\mathrm{3D}}$ whose evolution is governed by the interplay of large-scale gravitational and small-scale baryonic feedback processes. One way to probe the $\delta^{\mathrm{3D}}$ field is through gravitational lensing. Gravitational lensing is the bending of light rays coming from background light sources, e.g.\ galaxies, by the gravitational potential of foreground lens objects, e.g.\ galaxy clusters, resulting in our observation of shifted, magnified and distorted images. In most cases where a light ray passes far away from the centers of galaxy clusters where the gravitational potential is weak, it only experiences slight deflections by many foreground lenses distributed along its trajectory to us. Hence, we only observe a miniscule distortion in the image of the source. This is usually the case when light from sources pass through the foreground matter distribution of the large-scale structure (LSS). The distortion is so small that one can only see this effect on a statistical basis through correlations of the alignment of the weakly but coherently distorted images of many background source galaxies. This is known as \textit{weak gravitational lensing} and serves as a probe for investigating the distribution of matter in the LSS. This field can be interpreted as the \textit{shear} caused by a weighted line-of-sight projection of the 3D matter density contrast field --- known as the weak lensing convergence field. The weak lensing convergence field $\kappa(\boldsymbol{\theta})$ can be written as a line-of-sight projection of the 3D matter density contrast field \citep{Bartelmann_Schneider}
\begin{equation} \label{eq:convergence_definition}
\begin{split}
    \kappa(\boldsymbol{\theta}) & = \int \mathrm{d} \chi \; q(\chi) \delta^{\mathrm{3D}} \big(\chi\boldsymbol{\theta}, \tau_0 - \chi \big) \ ,
\end{split}
\end{equation}
where $\chi$ is the radial comoving distance, $\boldsymbol{\theta}$ is a 2D planar vector denoting positions on the sky and $\tau_0$ is the conformal time today. The term $q(\chi)$ is the projection kernel (also known as lensing efficiency) and for the case when all source galaxies are located in a Dirac-$\delta$ function like tomographic bin at $\chi_s$ it becomes: 
\begin{equation} \label{eq:lensing_projection_kernel_single_zs}
    q(\chi) = \frac{3H_0^2 \Omega_{\mathrm{m}}}{2c^2} \frac{\chi}{a(\chi)} \frac{\chi_s - \chi}{\chi_s} \; ;  \qquad \text{with } \chi \leq \chi_s \ .
\end{equation}
However, it is straight forward to write $q(\chi)$ for an ensemble of sources (instead of a single source) in terms of a distribution in a tomographic redshift bin following a normalized probability density function (PDF) $p(\chi')$ \citep{Kilbinger_2015}:
\begin{equation}
\label{eq:lensing_projection_kernel_ensemble_zs}
q_{\kappa}(\chi) = \frac{3H_0^2\Omega_{\mathrm{m}}}{2c^2}\frac{\chi}{a(\chi)} \int_\chi ^{\chi_{\rm lim}}{\rm d}\chi' p(\chi')\frac{\chi'-\chi}{\chi'} \; ;  \qquad \text{with } \chi \leq \chi_{\rm lim} \ .
\end{equation}
In the equations above $\Omega_{\mathrm{m}}$ is the total matter density parameter of the Universe today, $H_0$ the Hubble parameter today, $a$ the scale factor and $c$ the speed of light.

In this section we are interested in applying C3NN to a cosmological scenario where we study how well our architecture can distinguish simulated convergence fields which follow two different distributions: one a Gaussian random field and another a Log-normal random field. Log-normal random fields have been extensively studied and shown to closely approximate the 1-point PDF of the convergence field \citep{hilbert_lognormal_2011, xavier2016}. In order to simulate these sets of maps, we use the publicly available \verb|FLASK| tool \citep{xavier2016} to create realisations of Gaussian/Log-normal fields on the celestial sphere with a Dirac-$\delta$ source redshift distribution at $z = 1.0334$ in \verb|Healpix| format with \verb|NSIDE| = 2048 \citep{Gorski2005, Zonca2019}. The detailed process of creating the maps is described in section.~4.2 of \citealp{i3pcf_2021_halder}. The generated Gaussian and Log-normal maps share the same power spectrum (or the 2PCF) but differ in the higher-order correlations of the field which in the case of the Log-normal field is induced by the so called Log-normal shift parameter (see \citep{xavier2016}). This toy scenario is hence well setup for us to apply C3NN in a cosmological context and study its power in tapping into higher-order information in the lensing convergence field.

\begin{figure}[bt!]
\centering
\includegraphics[scale=0.6]{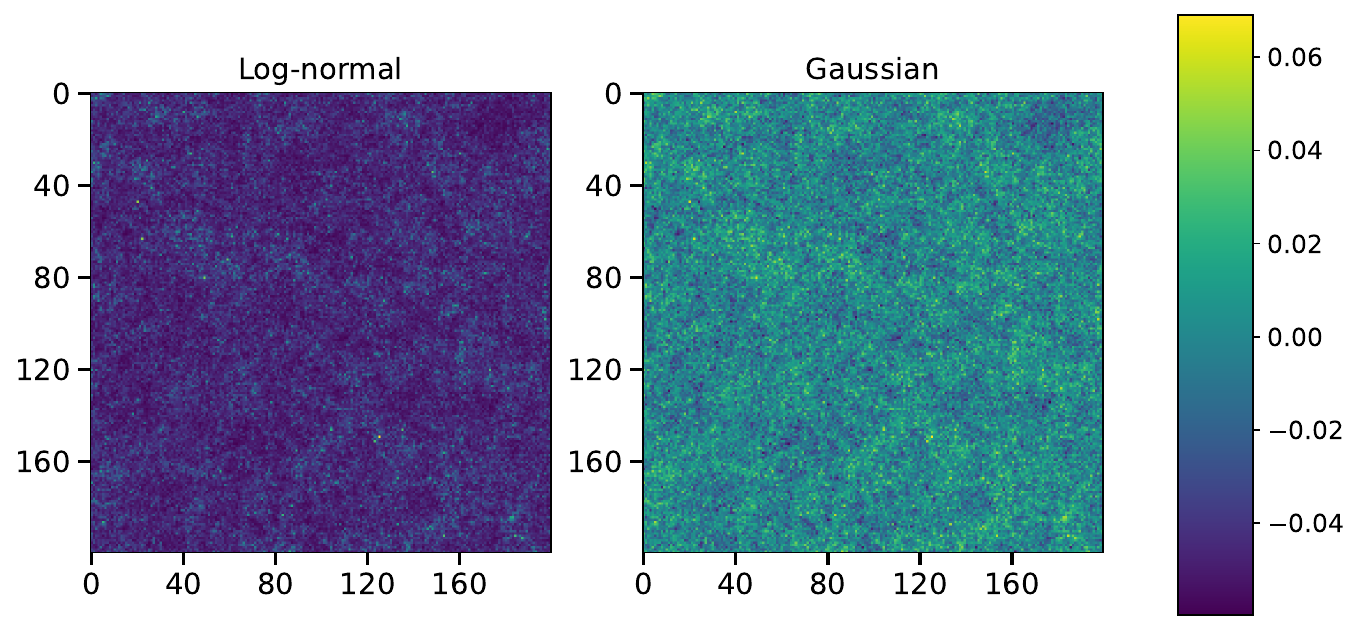}
\caption{Simulated Gaussian and Log-normal square maps with the same underlying cosmology and power spectrum but different higher-order correlation functions. Both classes of maps have size $200 \times 200$ pixels spanning an area of $20 \times 20$ $\rm deg^2$ on the spherical sky maps from which they are partitioned from (see Section \ref{sec:gaussian_lognormal_fields} for details). The color bar indicates the field amplitude.} 
\label{fig:grf_lognormal_flask}
\end{figure}

With the full-sky simulation maps from \verb|FLASK|, we partition them into non-overlapping square patches on which we can implement C3NN. We adopt the approach in \citet{Fulvio_MTNG_2023} (see their Section 2.4 and Figure 1) for partitioning the spherical sky. Each square map is $20 \times 20$ $\rm deg^2$ with an angular pixel resolution of 6 arcminutes. In Figure \ref{fig:grf_lognormal_flask} we show one example from Gaussian and Log-normal random field respectively. 
Both random fields have the same underlying cosmological parameters and power spectrum. As in Section \ref{sec:gaussian_fields}, we prepare 5000 maps for each class where 4500 is for training and 500 for validation. Each map only has 1 channel which in this case is a single source redshift bin lensing convergence field. We fix the filter number, filter size and the highest correlation order to be 1, $31 \times 31$ pixels and 4, respectively. The optimized hyper-parameters from \verb|Optuna| are shown in Table ~\ref{tab:ccnn_grf_lognormal}.
\begin{table}[bt!]
    \centering
    \begin{tabular}{|c|c|c|c|c|}
    \hline
      parameter & $\gamma$ & learning rate (lr) & learning rate decaying ratio ($\phi$) & optimizer \\
    \hline
      value & 0.0026 & 0.15 & 0.66 & ``RMSprop" \\
    \hline
    \end{tabular}
    \caption{Four optimized hyper-parameters in C3NN trained on Gaussian and Log-normal random fields (see Figure \ref{fig:grf_lognormal_flask}). $\gamma$ is the regularization strength in Eq.~\ref{eq:1st_training_L1_loss}. We use a learning rate scheduler which decays the initial learning rate (lr) of each parameter group by a factor $\phi$ after every epoch. We also simultaneously search for the best optimizer and obtain RMSprop \citep{rmsprop_2013}.}
    \label{tab:ccnn_grf_lognormal}
\end{table}
With these parameters, the model is able to reach a validation accuracy of $100\%$ after 100 training epochs.

\begin{figure}[bt!]
\centering
\includegraphics[width=0.4\linewidth]{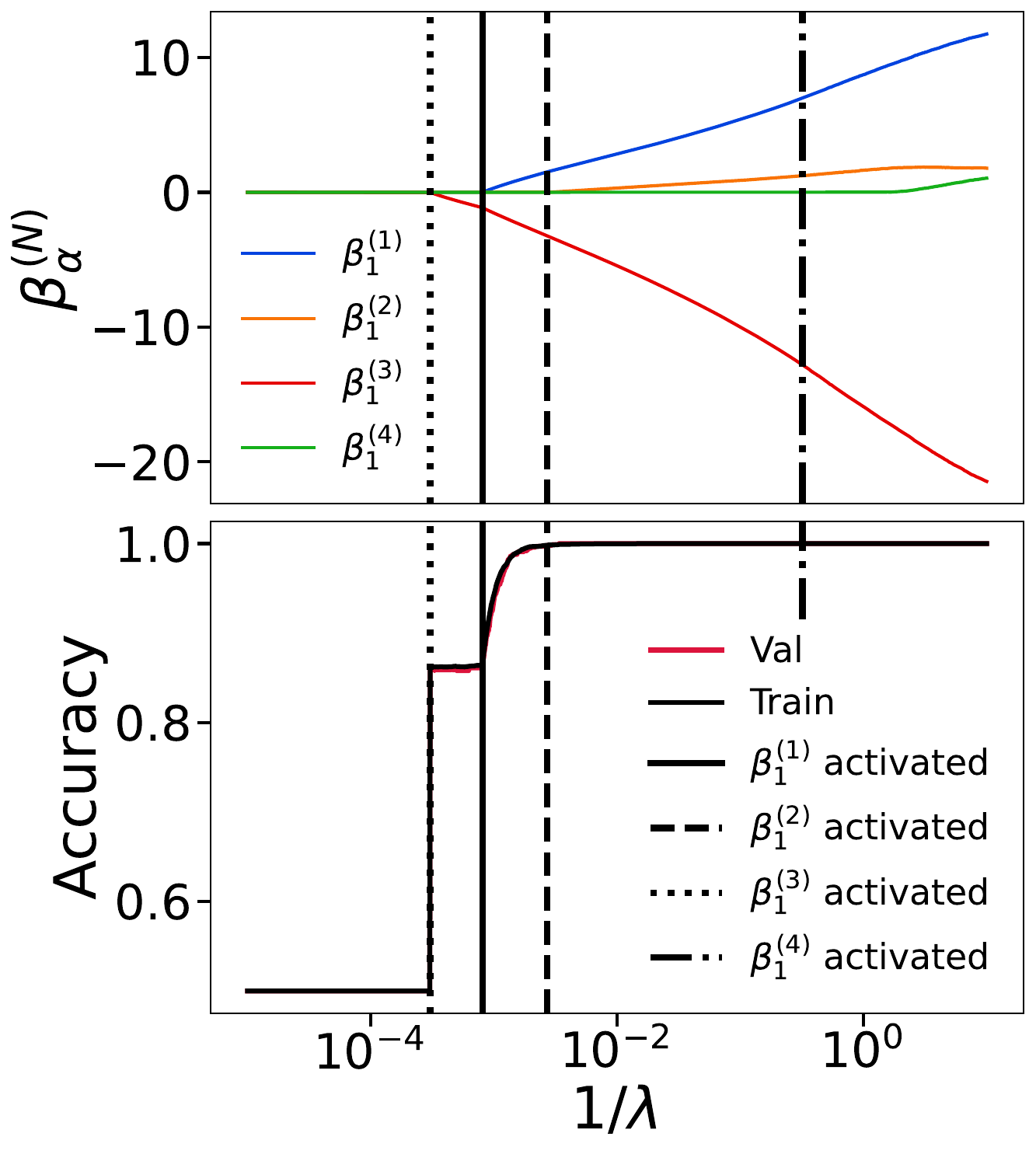}
\includegraphics[width=0.5\linewidth]{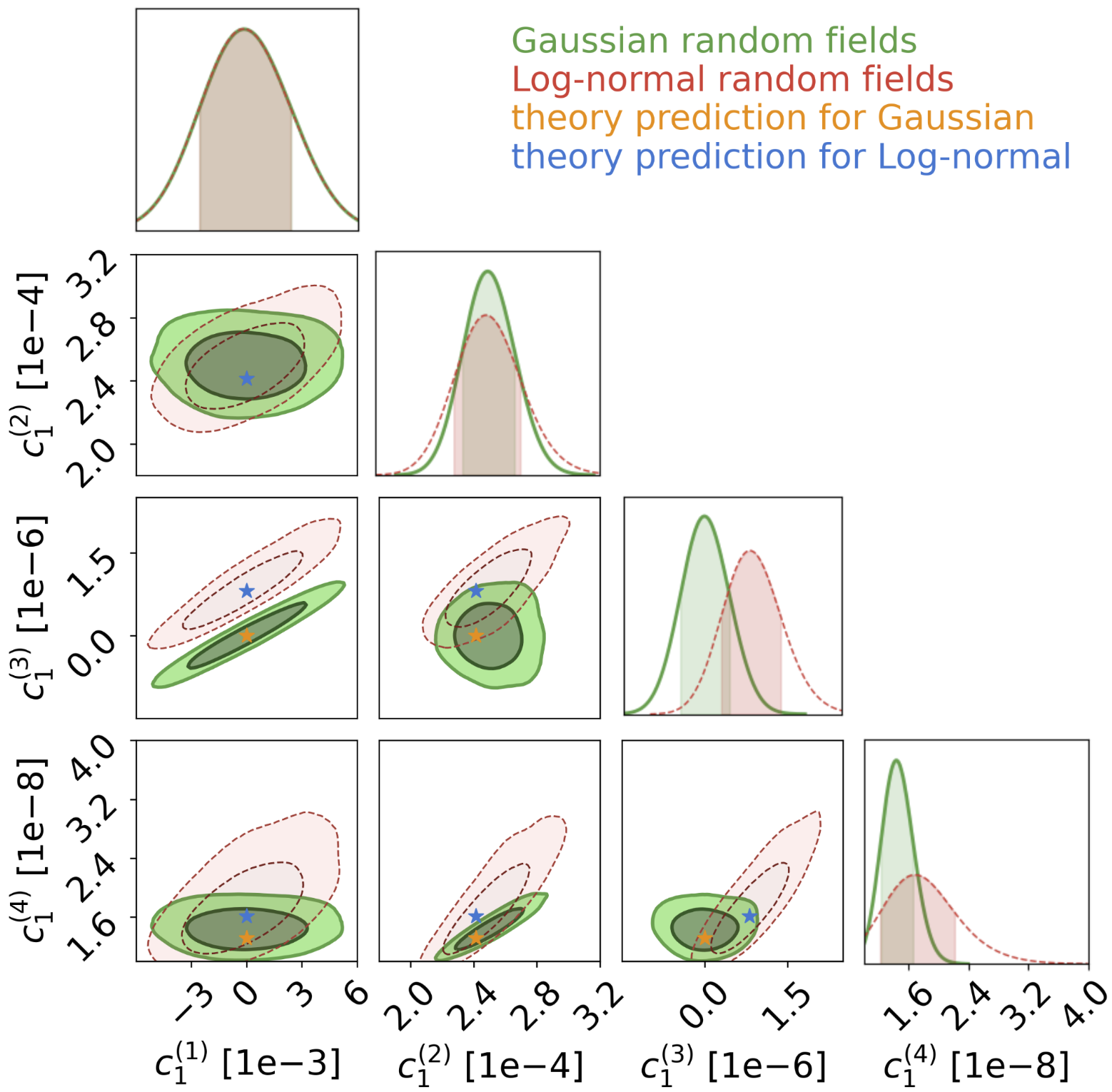}
\includegraphics[width=0.8\linewidth]{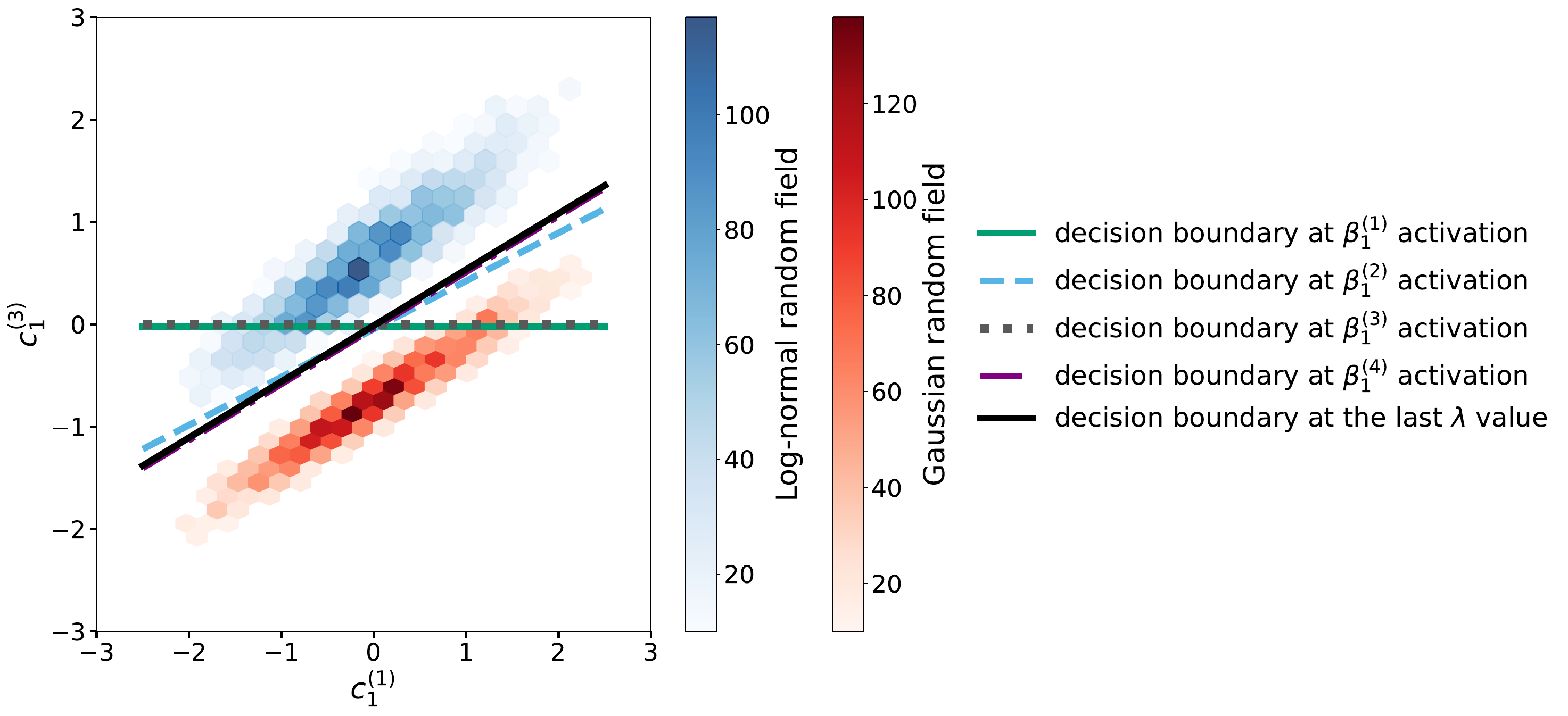}
\caption{\textit{Upper left}: The regularization path analysis applied to C3NN trained on two classes of Gaussian and Log-normal random fields (see Figure \ref{fig:grf_lognormal_flask}). The upper part shows the evolution of the $\beta_{\alpha}^{(N)}$ coefficients along with the decrease of the regularization strength $\lambda$ in Eq.~\ref{eq:2nd_training_L1_loss} (an increase in $1/\lambda$). The lower part shows the corresponding changes of training and validation accuracy. Different types of vertical dashed lines indicate the first $\lambda$ value at which each $\beta_{\alpha}^{(N)}$ becomes non-zero (i.e.~gets activated). \textit{Upper right}: Contours of moment $c_{\alpha}^{(N)}$ distributions mapped by passing the full training dataset into the moment map generator and spatial averaging after the first round of training. The diagonal subplots show the marginalized distributions of each order of moments from the two training classes. The green contours represent the moment distribution of the Gaussian random field and the red contours represent that of the Log-normal random field. The orange and blue stars are the theoretical predictions of different $c_{\alpha}^{(N)}$ for the two classes based on the trained filter weights and analytical expressions of correlation functions for the two fields. \textit{Bottom}: The distribution of $\{c_{1}^{(1)}, c_{1}^{(3)}\}$ with decision boundaries from the classifier. The values of both moments are batch normalized compared to the corresponding upper right panel. The blue distribution is from Log-normal random field and the red one is from Gaussian random field. Lines with different colors and styles corresponding to the vertical lines in the upper left panel represent the decision boundaries of C3NN at activations of $\beta_{\alpha}^{(N)}$ coefficients up to $N = 4$, as well as the boundary at the last regularization strength value.}
\label{fig:path_analysis_grf_lognormal}
\end{figure}

The regularization path analysis in the upper left panel of Figure \ref{fig:path_analysis_grf_lognormal} 
shows that the 3rd-order moment $c_{1}^{(3)}$ is activated first (dotted line) and contributes most to the classification accuracy. This is consistent with our expectation as the Gaussian and Log-normal random fields that we have constructed should only start to differ at the 3PCF level. The 2nd and 4th-order moments only become activated when both training and validation accuracy have reached $100\%$ so that their contributions can be counted as overfitting. One notices that $c_{1}^{(1)}$, which according to Eq.~\ref{eq:c_n_def} is the average of the field, is critical in complementing $c_{1}^{(3)}$ in the classification between the two classes. The reason behind this is explained in the bottom panel of Figure \ref{fig:path_analysis_grf_lognormal} where we show the marginalized distributions of feature vectors from training data in the $c_{1}^{(1)} - c_{1}^{(3)}$ plane. As described in Section \ref{sec:model_architecture}, we can visualize the evolution of the decision boundary in the projected $c_{1}^{(1)} - c_{1}^{(3)}$ plane, along with the changes of the regularization strength $\lambda$. At the very beginning of $\beta_{1}^{(3)}$ activation, the decision boundary can only be drawn based on the marginalized distribution of $c_{1}^{(3)}$ since all the other dimensions are still suppressed by the large $\lambda$ value. Once $\beta_{1}^{(1)}$ is activated, the decision boundary can be probed within a two-dimensional plane. It turns out that the mapped distributions of $\{c_{1}^{(1)}, c_{1}^{(3)}\}$ from the two random fields have the same degeneracy so that C3NN can rotate the decision boundary to achieve higher classification accuracy. This process is exactly depicted in the rotation from the green solid line ($\beta_{1}^{(1)}$ activation) to the blue dashed one ($\beta_{1}^{(2)}$ activation) in the figure. The blue decision boundary completely separates the two classes. Activation of $\beta_{1}^{(2)}$ together with others until the end of the regularization path makes almost no difference to the classification.  

As in the previous discussion, we can also combine the filter weights (see the right panel of Figure.~\ref{fig:trained_filter_grf_lognormal}) and the theoretical expressions for NPCFs to predict the corresponding moments $c_{1}^{(N)}$ for both the Gaussian and Log-normal fields. Any 2PCF of a given angular separation within the filter size is the same for both random fields and can be computed from the power spectrum. For the Gaussian random field all 4PCFs directly depend on the 2PCFs within the quadrilateral configurations while all its odd correlation functions are zero. For the Log-normal random field, any of its 3PCFs and 4PCFs can be written in terms of the 2PCF within a given configuration and the Log-normal shift parameter mentioned above. Readers who are interested in the exact equations are referred to Appendix B of \citet{hilbert_lognormal_2011}. The calculations give us $\{c_{1}^{(1)}, c_{1}^{(2)}, c_{1}^{(3)}, c_{1}^{(4)}\} = \{0.0, 2.414 \times 10^{-4}, 0.0, 1.315 \times 10^{-8}\}$ for the Gaussian random field and $\{c_{1}^{(1)}, c_{1}^{(2)}, c_{1}^{(3)}, c_{1}^{(4)}\} = \{0.0, 2.414 \times 10^{-4}, 8.078 \times 10^{-7}, 1.613 \times 10^{-8}\}$ for the Log-normal random field. These are marked by the orange and blue stars in the upper right panel of Figure \ref{fig:path_analysis_grf_lognormal} respectively. Once again, the calculations fall well within the bulk of the moments distribution output by C3NN which we obtain by passing all the training dataset through our trained model. The right panel of Figure \ref{fig:trained_filter_grf_lognormal} shows that the dominant weights that hold the classification power mainly concentrate in the central part of the filter. Qualitatively, we can understand this as statistically speaking, Log-normal random fields asymptotically resembles Gaussian random fields when smoothed on large scales. The differences between the two fields are mainly captured through the higher-order correlation functions which become increasingly more significant on small scales. An efficient way to distinguish the two classes therefore would be to select correlation functions beyond 2PCFs on small scales. Then it is no wonder that out of all triangular configurations that can be selected from the filter constituting $c_{1}^{(3)}$, C3NN considers the smallest one the most important in aiding the classification which is an isosceles triangle with two side lengths equal to 6 arcmin (neighboring pixel separation) and the third equal to $6\sqrt{2}$ arcmin. 

\section{Results from N-body cosmological simulations}
\label{sec:n_body_results}
In Section \ref{sec:gaussian_fields} and \ref{sec:gaussian_lognormal_fields}, we have shown that C3NN can classify different random fields by correctly capturing the underlying statistical properties. In this section we are interested in applying the framework to weak lensing fields from realistic N-body simulations to investigate whether a successful classification is still possible and what statistical features C3NN extracts from the training data. We shall also investigate the impact of shape noise and smoothing scale of the simulated maps on the model's performance.

\subsection{Weak lensing convergence maps from N-body simulations}
\label{sec:simulations}
In the rest of this work, we use weak lensing convergence maps from the publicly available \textit{CosmoGridV1}\footnote{www.cosmogrid.ai} simulations \citep{fluri_kids1000, cosmogridv1_2023}. We focus on the simulation products around the fiducial cosmology adopted by \textit{CosmoGridV1} which has the following parameter values: $\Omega_{\rm m}=0.26$, $\Omega_{\rm b}=0.0493$, $\sigma_8=0.84$, $w_0=-1.0$, $n_{\rm s}=0.9649$ and $H_0=67.3$ $\rm km/s/Mpc$. The two classes of maps we use as training data share the same parameters as the fiducial one except for $w_0$. One class of simulation has $w_0=-1.05$ which is in the realm of ``phantom dark energy" \citep{phantom_de_review_2017} and the other has $w_0=-0.95$ which is characterized as ``quintessence" \citep{quintessence_review}. For each simulation set there are 200 independent full-sky simulations. Each simulation contains a series of lightcone shells along the redshift until $z=3.5$. We project these shells following the four source galaxy redshift distributions of the Dark Energy Survey Year 3 (DES Y3) data release (see Figure 11 in \citet{Myles_2021}) to produce the corresponding full-sky weak lensing convergence maps. Readers can refer to \citet{ufalcon_1,ufalcon_2} and \citet{ufalcon_3} for details of the methodology we adopt for the shell projection and convergence map generation. These maps from the four different source redshift distributions constitute the four channels in the input data and enable us to perform a tomographic analysis with C3NN. 

On top of these noiseless maps, we also add shape noise in order to make more realistic tests. We directly add noise to each pixel in the full-sky convergence map by sampling independently from a Gaussian distribution with zero mean and variance $\sigma^2 = \epsilon^2/({\rm A}n_{\rm g})$ where $\epsilon$ is the dispersion of galaxy ellipticity, $n_{\rm g}$ is the galaxy surface number density and $\rm A$ is the pixel area. We adopt DES Y3 values for $\epsilon$ and  $n_{\rm g}$ which are $0.255$, and $\{1.476, 1.479, 1.484, 1.461\}$ ${\rm arcmin^{-2}}$ for each source redshift bin respectively \citep{desy3_massmap_2021}. The pixel area of the full-sky convergence map is 2.95 $\rm arcmin^2$. Besides shape noise we also take into account the smoothing of the input maps. This is motivated by the concern that in real observation we may need to discard small scale measurement of correlation functions since they are impacted by the baryonic feedback effects that we cannot realistically model \citep{des_y3_3_cross_2pt_2022} in N-body gravity only simulations. For map smoothing, we apply Gaussian kernels with full widths at half maximum (FWHM) of $\{10^{\prime}, 20^{\prime}, 30^{\prime}, 40^{\prime}\}$ to the tomographic full-sky maps.

In this analysis, we use the same partition method mentioned in the previous section to obtain 5800 non-overlapping square maps for each cosmology and source redshift bin. We split $90\%$ of the data for training and $10\%$ for validation. Regardless of the smoothing scales, the square map always has an area of $20 \times 20$ $\rm deg^2$ and a size of $200 \times 200$ pixels. The difference of $w_0$ values between the two cosmologies is about $1/6$ of the marginalized $1\sigma$ uncertainty as constrained by the state-of-the-art DES Y3 analysis \citep{des_y3_3_cross_2pt_2022}. This constraint comes from a combination of cosmic shear and galaxy clustering, the so-called $3 \times \rm 2pt$ analysis, within the $w_0\rm CDM$ model. Only when data from $3 \times \rm 2pt$ and the external low-redshift probes (including Baryon Acoustic Oscillation (BAO), Supernovae Type Ia (SNe Ia) and Redshift Space Distortion (RSD)) are combined can the marginalized $1\sigma$ uncertainty of $w_0$ be constrained approximately to 0.1. Based on these contexts, it would require multiple powerful cosmological probes to accurately distinguish between the two dark energy scenarios we are testing. Hence, in the context of the above training setups, we would like to address the following questions: (i) with weak lensing convergence alone (i.e.~no external probes), how much generic improvement can higher-order correlation functions as captured by the moments $c_{\alpha}^{(N)}$ contribute to differentiating these two dark energy equation of state parameter; (ii) what is the relative importance of each order of moment in such a classification task; (iii) how does shape noise and different smoothing scales affect the model performance.

\subsection{Noiseless tomographic training results with varying smoothing scales}
\label{sec:tomographic_analysis_noiseless}
\begin{figure}[bt!]
    \centering
    \includegraphics[height=8cm]{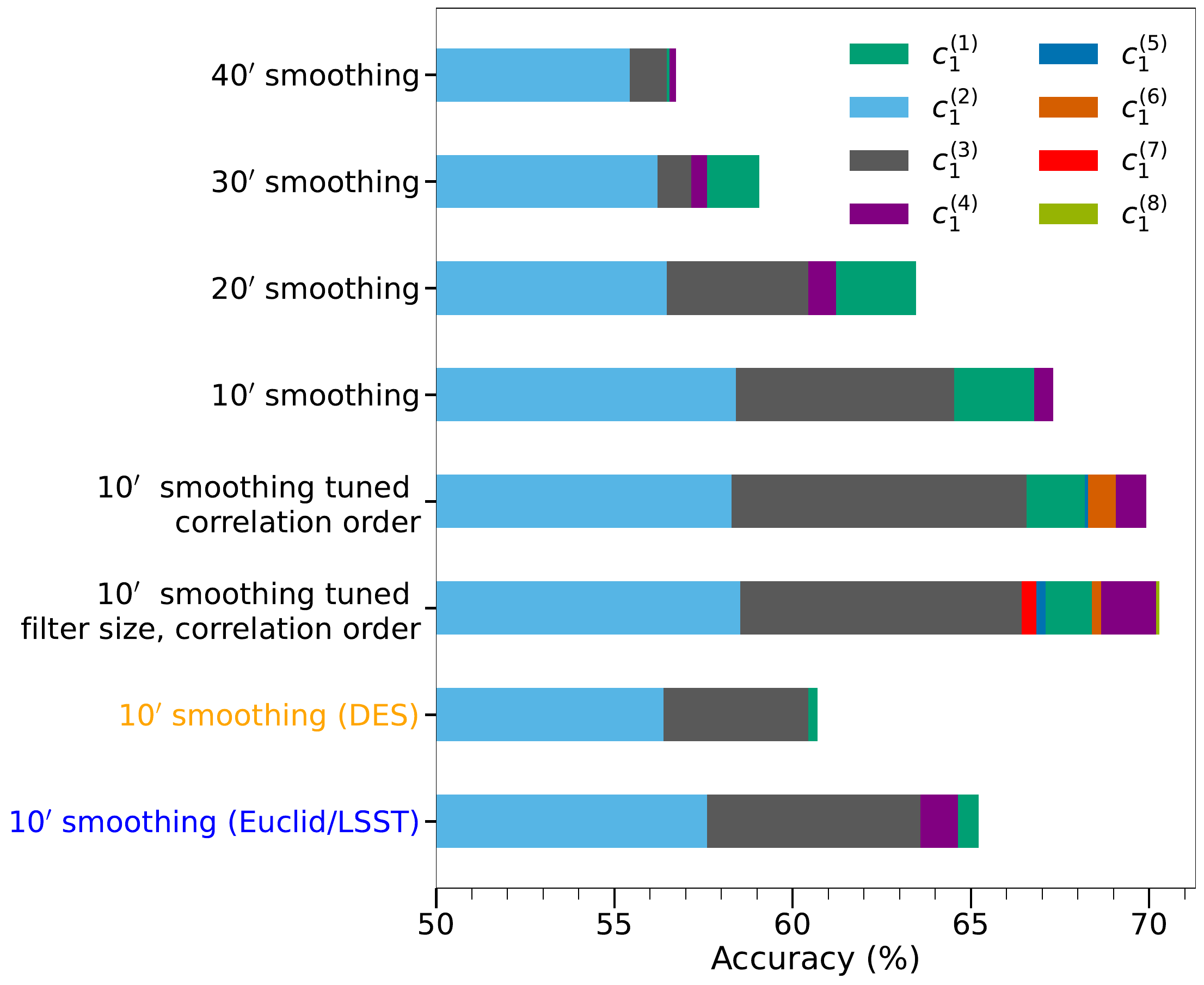}
    \includegraphics[height=8cm]{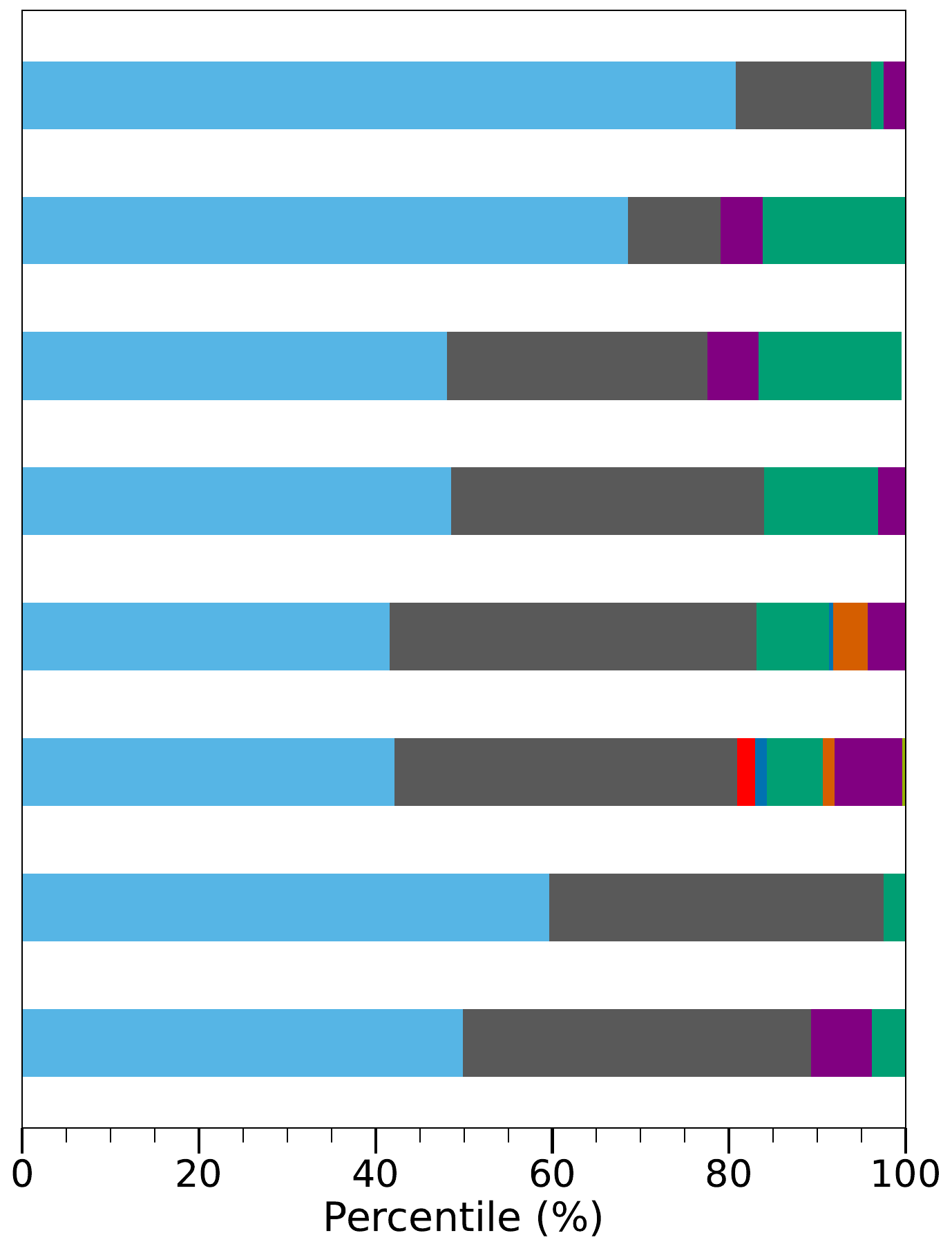}
    \caption{\textit{Left}: The validation accuracy from regularization path analysis (see Section \ref{sec:model_architecture}) of C3NN trained on two classes of tomographic weak lensing convergence maps with different $w_0$ values (-1.05 and -0.95). The four source redshift distributions follow that of DES Y3 analyses. Test names in black indicate that the analyses are with noiseless data. The first four tests deviate from each in the FWHM of the Gaussian kernels used for smoothing. In the 5th row of the bar chart, we give the model one extra free parameter by allowing \texttt{Optuna} to optimize the highest correlation order while still fixing the filter number and size. In the 6th row, we further add filter size to the set of tuning parameters. The sixth and seventh rows of the bar charts are results from different noisy training data with the same model setups as the corresponding noiseless test with 10' smoothing. The name in orange represents that the test is with noisy data produced with DES Y3 $\epsilon$ and $n_g$ values while the one in blue uses training data with $n_g$ rescaled to 7.5 $\rm arcmin^{-2}$ for each tomographic bin (as expected for a Stage-IV survey like Euclid/LSST). In each single row of the bar chart, from left to right, the different color components form the sequence of the activation of moments in the regularization path analysis. The length of every color component represents the improvement in validation accuracy added by the current activation combined with all the previous ones. The total validation accuracy starts from $50\%$ (random classification) and ends with the summation of all color components. \textit{Right}: The percentage of improved validation accuracy from the activation of different moments with respect to the total validation accuracy in each test. It shares the same color labels and test names as the left panel. Color components in each bar are also arranged according to the activation sequence.}
    \label{fig:ccnn_w0_noiseless_noisy}
\end{figure}

With the above mentioned noiseless training data, we apply C3NN to them at each smoothing scale and summarize the results in the left panel of Figure \ref{fig:ccnn_w0_noiseless_noisy}. To keep the model simple, at the beginning we freeze the filter number, filter size and the highest correlation order (1, $31 \times 31$ pixels and 4 respectively) while changing the smoothing scales of the training data. The tuned hyper-parameters are shown in Table \ref{tab:ccnn_noiseless_w0_nofree_param}.
\begin{table}[bt!]
    \centering
    \begin{tabular}{|c|c|c|c|c|}
    \hline
      & $\gamma$ & learning rate (lr) & learning rate decaying ratio ($\phi$) & optimizer \\
    \hline
      $40^{\prime}$ $\rm smoothing$ & 0.06 & 0.018 & 0.58 & ``Adam" \\
      $30^{\prime}$ $\rm smoothing$ & 1.10 & 0.041 & 0.24 & ``Adam" \\
      $20^{\prime}$ $\rm smoothing$ & 0.0023 & 0.043 & 0.34 & ``Adam" \\
      $10^{\prime}$ $\rm smoothing$ & 0.0022 & 0.064 & 0.14 & ``Adam" \\
    \hline
    \end{tabular}
    \caption{Same as Table \ref{tab:ccnn_grf}, the optimized hyper-parameters of C3NN for varying smoothing scales applied to noiseless tomographic weak lensing convergence maps with different $w_0$ values (-1.05 and -0.95) (see Section \ref{sec:tomographic_analysis_noiseless}). Here we fix the filter number, filter size and the highest correlation order.}
    \label{tab:ccnn_noiseless_w0_nofree_param}
\end{table}

From the figure we can observe that the 2nd and 3rd-order moments are the major contributors to the classification. The activation of $\beta_1^{(3)}$ (onset of the grey bars) always follows that of $\beta_1^{(2)}$ (end of the light blue bars), indicating that when one only searches for a single moment to distinguish the above two $w_0\rm CDM$ cosmologies, the 2nd-order moment $c_1^{(2)}$ is most effective. However, once the 3rd-order moment $c_1^{(3)}$ is activated and combined with $c_1^{(2)}$, it can significantly improve the validation accuracy. This is analogous to what \citet{i3pcf_gong_2023} found from the perspective of parameter inference that $w_0$ constraint can be particularly tightened by combining 2PCF and a compressed version of the 3PCF (called the integrated 3PCF). As for the other higher-order moments, here $c_1^{(4)}$ specifically, can only bring minor validation accuracy improvement compared to $c_1^{(2)}$ and $c_1^{(3)}$. This can be explained by either the cosmic variance caused by the lack of sampling of 4PCF configurations within the filter due to the limited 400 $\rm deg^2$ simulated map area resulting in a lower signal-to-noise compared to lower order moments, or the map resolution which prohibits the higher-order moments in probing smaller and smaller scales where their importance become increasingly more significant. Both effects can lead to the weakening of the classification capability of higher-order moments. Another explanation for the weak classification power contribution from $c_1^{(4)}$ under the current setups is that there is not much information contained in the 4th-order moment of weak lensing convergence that can be added to differentiate between these two different cosmologies. One supportive evidence is that the increased validation accuracy after activating $\beta_{1}^{(4)}$ is not as significant as the increment brought by $c_1^{(3)}$.  

\begin{figure}[bt!]
\centering
\includegraphics[height=6cm]{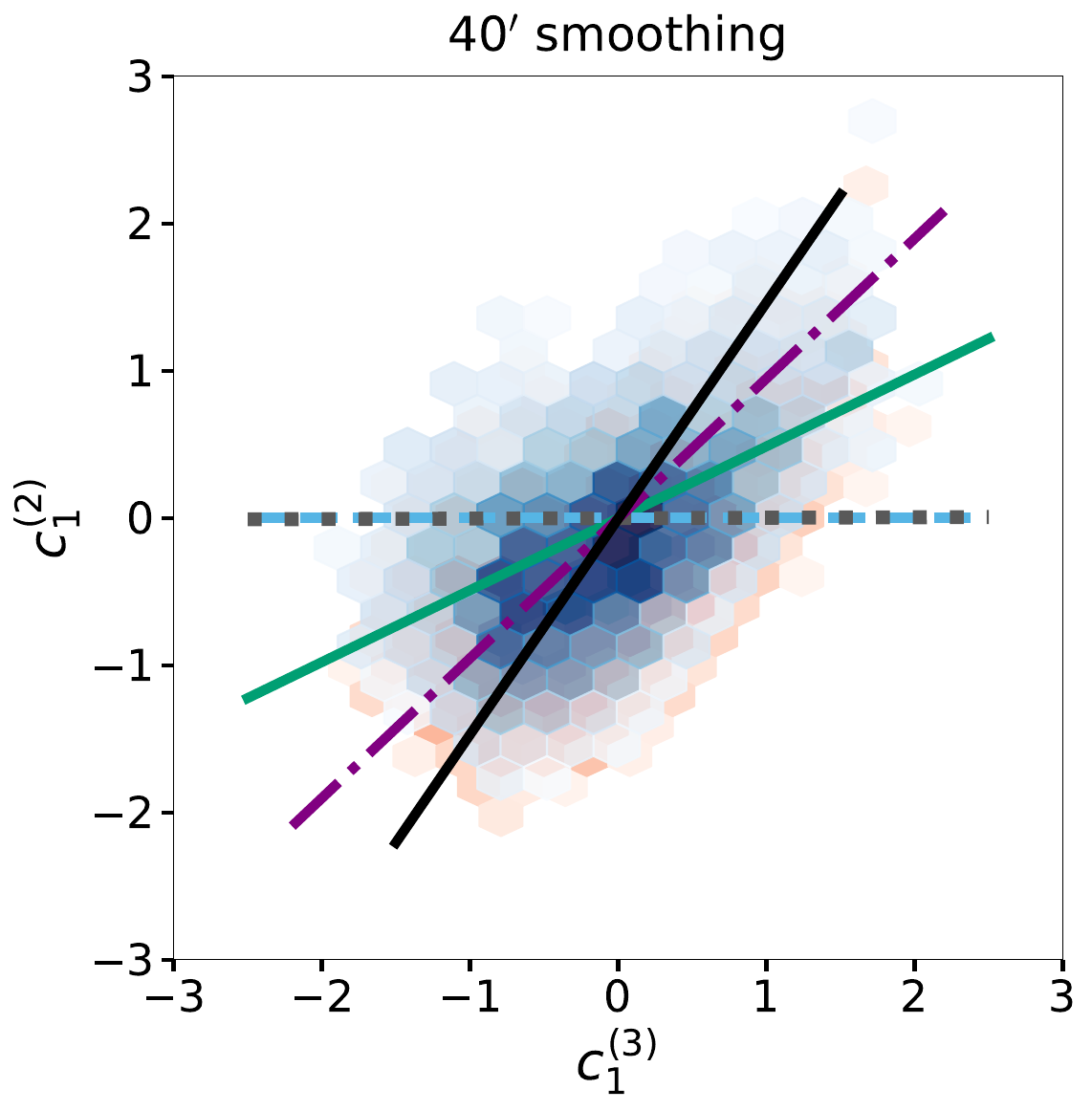}
\includegraphics[height=6cm]{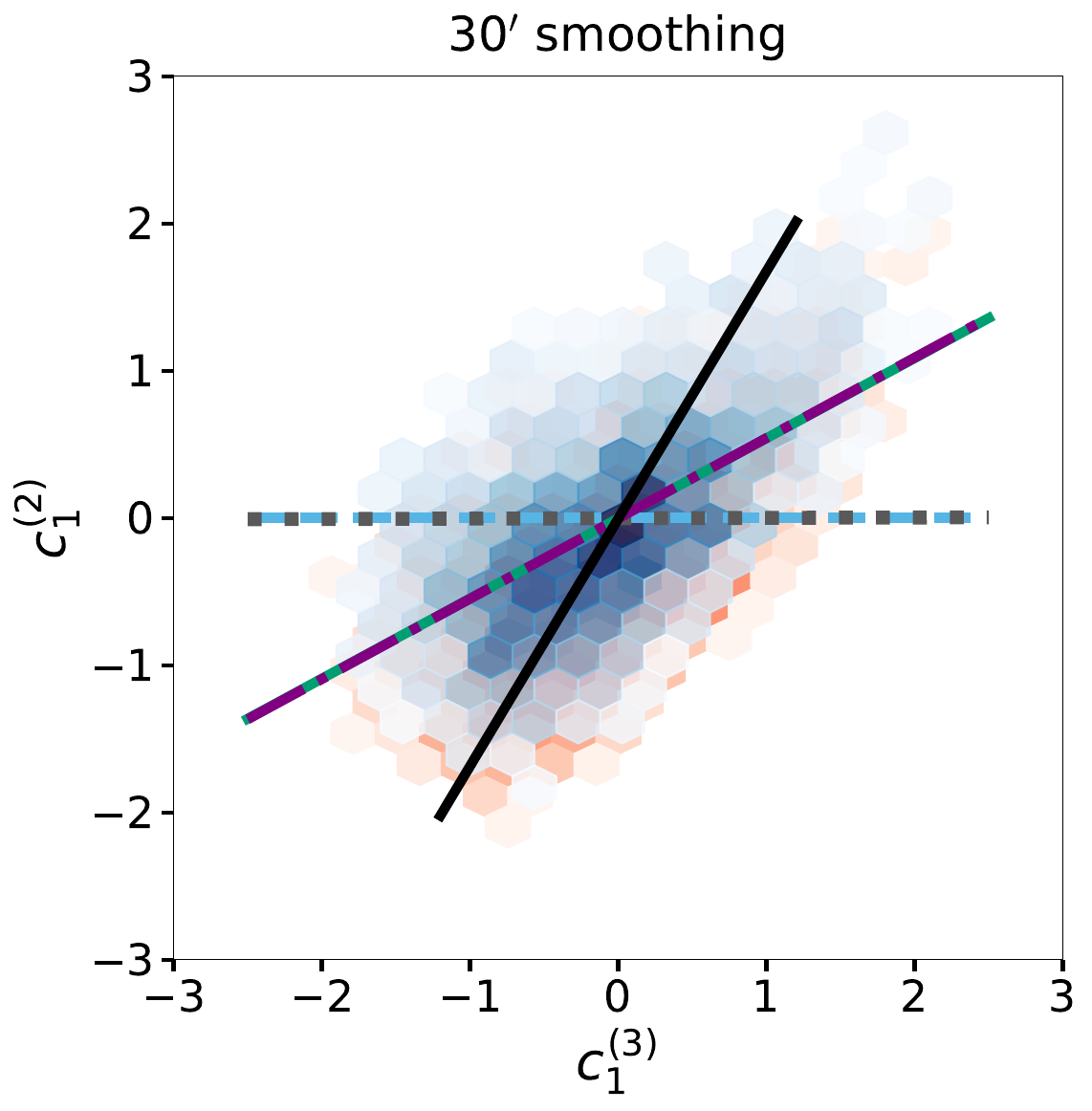}
\includegraphics[height=6cm]{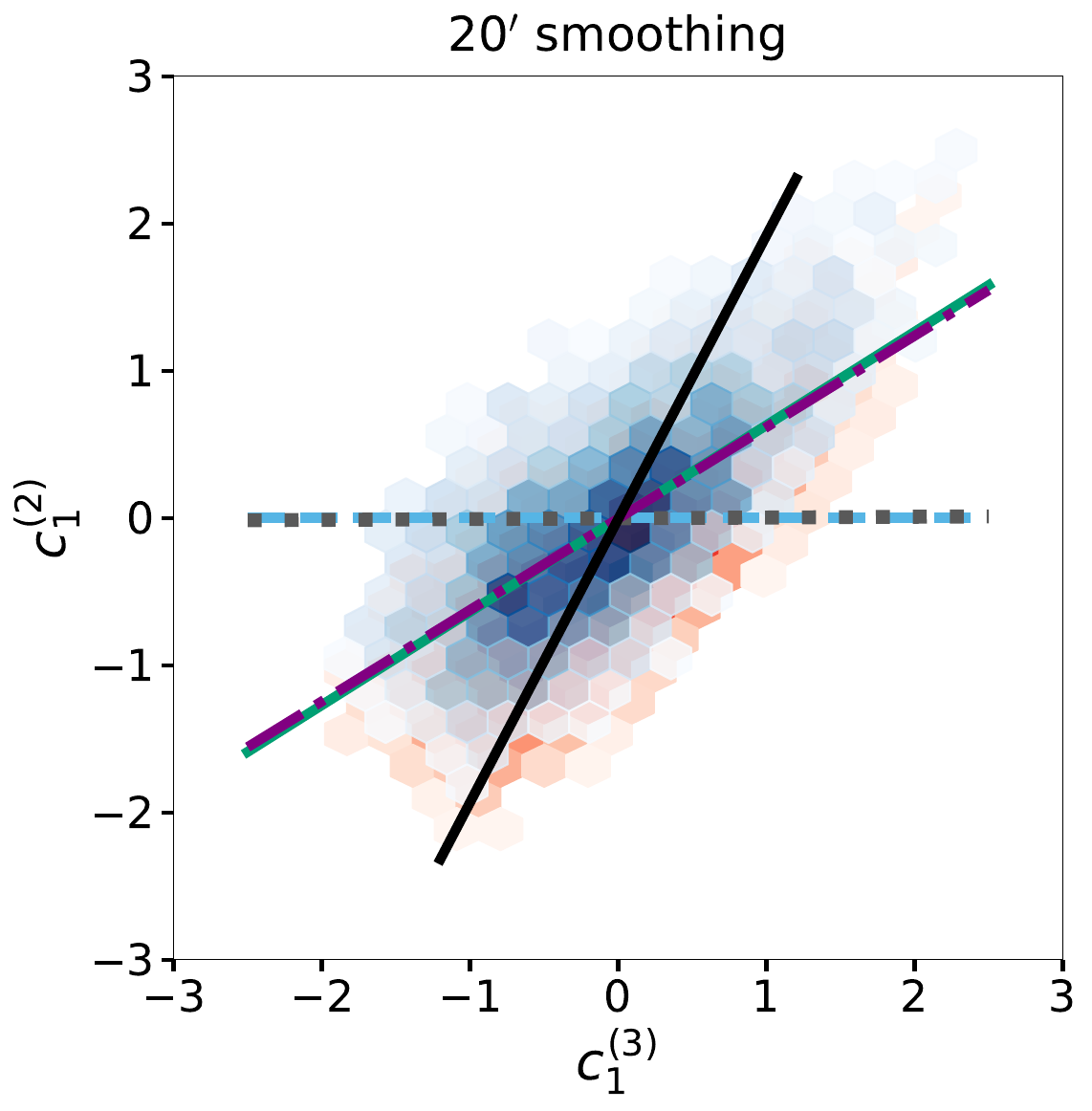}
\includegraphics[height=6cm]{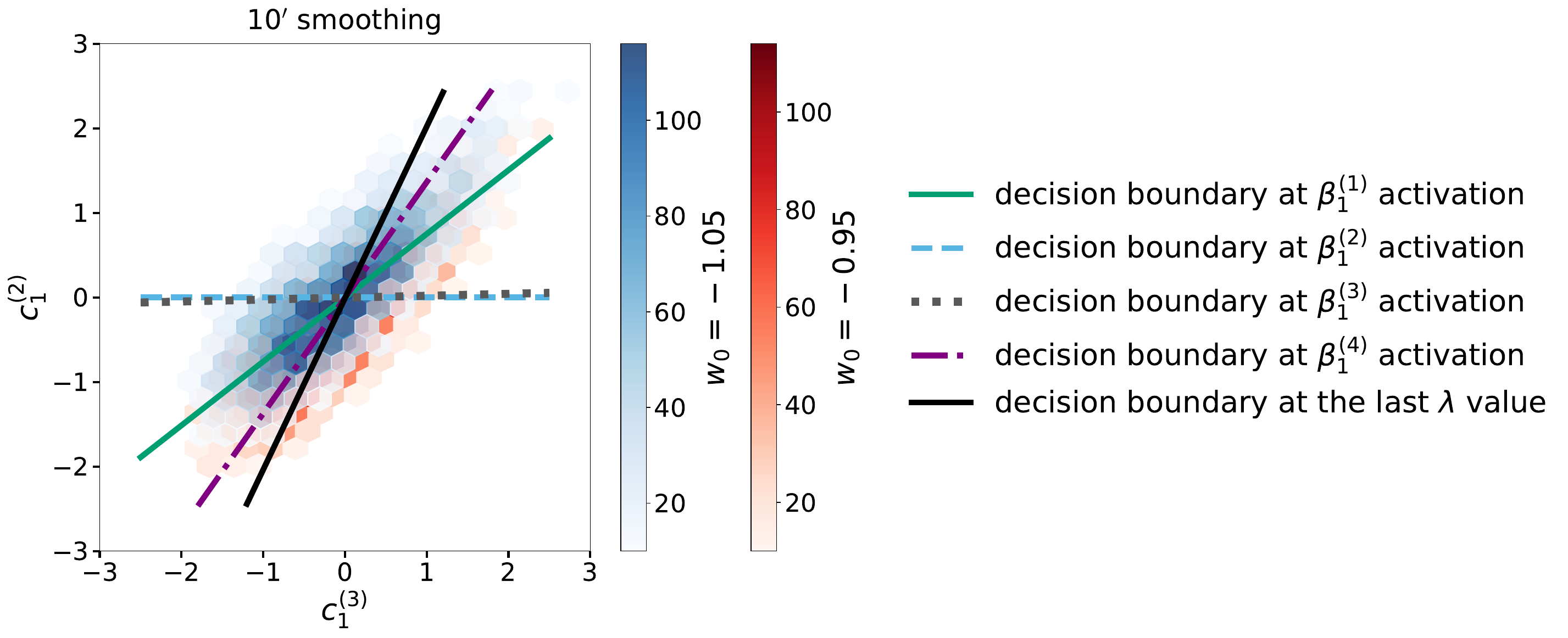}
\caption{The batch normalized $\{c_1^{(3)}, c_1^{(2)}\}$ joint distributions mapped by C3NN trained on simulated noiseless weak lensing convergence data for different smoothing scales. The blue distribution represents the mapping from data with $w_0=-1.05$ and the red distribution represents the mapping from data with $w_0=-0.95$. Lines with different colors and styles are the decision boundaries learned by the classifier during the regularization path analysis at the activation of different orders of moments.}
\label{fig:noiseless_w0_c2_c3_distribution}
\end{figure}

One should also notice that although the amount of validation accuracy associated with $c_1^{(2)}$ (i.e. the size of the light blue bars in the first four rows in the left panel of Figure \ref{fig:ccnn_w0_noiseless_noisy}) increases along with the decrease of the smoothing scale, the results are still similar to each other i.e.~they make up ~$6\% - 8\%$ of the total validation accuracy. In comparison, the additional validation accuracy brought by $c_1^{(3)}$ (size of the grey bars) depends more drastically on the smoothing scale. The result from $40^{\prime}$ smoothing is approximately 6 times smaller than the result from $10^{\prime}$ smoothing. This different dependence indicates that the joint distribution of $\{c_1^{(2)}, c_1^{(3)}\}$ mapped by the trained model from two classes has larger overlapping region as we increase the smoothing scale. Similar to the bottom panel in Figure \ref{fig:path_analysis_grf_lognormal}, we show in Figure \ref{fig:noiseless_w0_c2_c3_distribution} the batch-normalized $\{c_1^{(2)}, c_1^{(3)}\}$ joint distributions for the different smoothing scales.
It is clear that the initial activation of $\beta_1^{(2)}$ leads to the horizontal decision boundary in each panel. Those decision boundaries then gradually rotate to the direction along the degeneracy of $c_1^{(2)}$ and $c_1^{(3)}$ after other activations. However, as the smoothing scale increases, the distributions from two classes in $c_1^{(2)}-c_1^{(3)}$ plane become more extended along the orthogonal direction of the degeneracy. This naturally enlarges the overlapping region of the two distributions and reduces the validation accuracy C3NN can obtain with the activation of $\beta_1^{(3)}$.

In order to ensure that our validation accuracy is not limited by the fixed model parameters, i.e.~filter number, filter size and the highest correlation order, we alternatively treat them as hyper-parameters and optimize them using \verb|Optuna|. The 5th and 6th rows of the bar charts in the left panel of Figure \ref{fig:ccnn_w0_noiseless_noisy} show the results of tuning the highest correlation order and simultaneously tuning it with the filter size respectively (see Table~\ref{tab:ccnn_noiseless_w0_free_corder} and \ref{tab:ccnn_noiseless_w0_free_size_and_corder} for the results). 
\begin{table}[bt!]
    \centering
    \begin{tabular}{|c|c|c|c|c|c|}
        \hline
        parameter & correlation order & $\gamma$ & learning rate (lr) & learning rate decaying ratio ($\phi$) & optimizer \\
        \hline
        value & 6 & 0.0039 & 0.18 & 0.75 & ``Adam" \\
        \hline
    \end{tabular}
    \caption{Optimized hyper-parameters including the highest correlation order for the C3NN model trained on simulated noiseless tomographic convergence maps with $10^{\prime}$ smoothing (see Section \ref{sec:tomographic_analysis_noiseless}).}
    \label{tab:ccnn_noiseless_w0_free_corder}
\end{table}
In both cases where C3NN can measure moments beyond 4th order, the validation accuracy only shows little improvement. The dominant contribution still comes from the joint distribution of $\{c_1^{(1)}, c_1^{(2)}, c_1^{(3)}\}$. We even further allow the tuning process to vary the number of filters in the model but it does not help to substantially increase the validation accuracy either. We do not show the corresponding result in the plot as the optimized filter number is 5 and the highest correlation order is 8 according to \verb|Optuna| which is difficult to fit into the plotting space. This is another sign suggesting that moments of convergence beyond 3rd order in total may not contain sufficient information to classify different dark energy equation of state parameters at the desired precision of $\Delta w_0 = 0.1$ with our test setups. To address this issue, one promising approach would be to include more observables such as projected galaxy number density into the training data and measure the cross-correlations among these quantities rather than keep adding higher-order moments beyond 3rd order in the data vector. We defer this study to future works.
\begin{table}[bt!]
    \centering
    \begin{tabular}{|c|c|c|c|c|c|c|}
        \hline
        parameter & filter size & correlation order & $\gamma$ & learning rate (lr) & learning rate decaying ratio ($\phi$) & optimizer \\
        \hline
        value & 11 $\times$ 11 & 8 & 0.002 & 0.019 & 0.93 & ``Adam" \\
        \hline
    \end{tabular}
    \caption{Same as Table \ref{tab:ccnn_noiseless_w0_free_corder} but also allowing the optimization of the filter size along with the highest correlation order in the training.}
    \label{tab:ccnn_noiseless_w0_free_size_and_corder}
\end{table}

From the above discussion, we show that C3NN can be a useful tool to investigate the sensitivity of moments with respect to $w_0$ parameter within $w_0\rm CDM$ model. Broadly speaking, it is possible to extend this functionality to any given binary classification tasks of two cosmological models that can impact differently the matter distribution in our Universe, e.g. between the standard cold dark matter and other exotic models such as wave dark matter \citep{wave_dm_review} or self-interacting dark matter \citep{sidm_review}. Moreover, it can provide a quantitative understanding of the relative importance of different orders of moments, which are directly related to correlation functions, in classifying different cosmological parameters or models. All exact numeric values in Figure \ref{fig:ccnn_w0_noiseless_noisy} may vary slightly from one evaluation to another, but the comparison among the total validation accuracy of different tests is certain. Moreover can always confirm the dominance of 2nd and 3rd-order moments as the first and second activated features carrying relatively the most significant information in the classification.

\subsection{Noisy tomographic training results}
\label{sec:tomographic_analysis_noisy}
For simplicity, we only show the training result from data at $10^{\prime}$ smoothing scale with shape noise with DES Y3 dispersion of galaxy ellipticity and surface number density. This is the 7th row of bar chart in the left panel of Figure \ref{fig:ccnn_w0_noiseless_noisy}. Validation accuracy from the rest smoothing scales with the same noise addition gradually decreases as in the noiseless cases. Compared to the corresponding noiseless result (4th row), here the total validation accuracy reduces by approximately $6.6\%$ and $c_1^{(3)}$ contributes less significantly to the classification power. Moreover, the contributions from both $\beta_1^{(1)}$ and $\beta_1^{(4)}$ almost vanish in the activation sequence. Overall, the effects of shape noise are similar to that of smoothing: it mainly decreases the total validation accuracy through moments beyond 2nd order.

Besides adding the shape noise associated with DES Y3 analyses, we also rescale the galaxy surface number density mentioned in Section \ref{sec:simulations} to approximately the value that would be provided by upcoming Stage-IV lensing surveys such as Euclid \citep{Euclid_preparation} and Vera Rubin’s LSST \citep{LSST_survey}. We adopt the new total galaxy surface number density to be 28 galaxies per $\rm arcmin^{2}$ which is between those of the two surveys \citep{Euclid_shape_noise,LSST_shape_noise}. We retain the four tomographic bins in the training data and equally divide this new number density among all of them. The original shape noise thus should be rescaled by a factor of $\{0.444, 0.444, 0.445, 0.441\}$ for each tomographic bin respectively. Through this we obtain a crude estimation of C3NN performance on the next generation survey data. One should bear in mind that in this test the map size of the training data ($20 \times 20$ $\rm deg^2$) is much smaller than the actual footprints of these Stage-IV surveys and that we also use less number of tomographic bins. The result is shown in the 8th row of the bar chart in the left panel of Figure \ref{fig:ccnn_w0_noiseless_noisy}. Clearly with a higher signal-to-noise ratio, the total validation accuracy exceeds that of noisy simulated data using DES Y3 shape noise parameters. Out of all moments, $c_1^{(3)}$ gains the largest relative growth in its classification power. Although relatively small compared to $c_1^{(2)}$ and $c_1^{(3)}$, we also find that $c_1^{(1)}$ and $c_1^{(4)}$ can bring non-negligible improvements to the total validation accuracy. Another point to observe is that though it is not as powerful as the model trained on the corresponding noiseless data as expected, there is not a great contrast between the results that these two models manifest. This implies that with the forthcoming Stage-IV surveys, C3NN can potentially be close to reaching its best performance.

Another useful perspective to understand all the above results is shown in the right panel of Figure \ref{fig:ccnn_w0_noiseless_noisy}. The total validation accuracy of each test is used as a normalization constant, based on which we calculate the percentage of increased validation accuracy associated with each order moment. We notice that with larger smoothing scale or shape noise, the percentage of the 2nd-order moment increases while that of higher orders, here particularly 3rd and 4th-order, decreases. This is consistent with our previous findings that the power of 2nd-order moment is robust against varying smoothing scales or shape noise but higher-order moments can be significantly impacted. Another point the right panel helps to stress is that $c_1^{(3)}$ is powerfully complementary to $c_1^{(2)}$, e.g.~in the test case of noisy data with Euclid/LSST galaxy surface number density, the combination of $c_1^{(3)}$ and $c_1^{(2)}$ almost double the classification ability of the model compared to $c_1^{(2)}$ alone. Unfortunately we do not observe this feature in moments beyond 3rd order within our test setups. 

\section{Summary and conclusion}
\label{sec:summary_and_conclusion}
C3NN, originally proposed by \citet{Miles_ccnn_2021} in the context of correlated quantum matter, is an interpretable machine learning architecture that we have introduced for cosmological analyses. It is composed of two parts (see Figure \ref{fig:ccnn_architecture}). The first part comprises a CNN based N-point moment map generator which outputs a series of statistics $c_{\alpha}^{(N)}$ which we name as N-point moments and have a one-to-one correspondence with traditional NPCFs that we are familiar with in cosmology. In this part of the model we first perform a single convolution on the input data and then without applying a nonlinear transformation as the usual practice, we construct moment maps directly from the initial convolved map through a recursive procedure (see Eq.~\eqref{eq:C_N_recursive}). We adapt the filters of the convolution to be rotationally invariant \citep{escnn_cesa_2022} such that the output remains stationary with respect to arbitrary rotations of the input. This benefits the training data efficiency of our analyses and the interpretation of the trained filter weights. The second part of the model which is a moment-map based classifier exploits these output moment maps to perform classification tasks by passing them through consecutive layers of spatial averaging, batch normalization and logistic regression. We optimize all hyper-parameters of C3NN using the package \texttt{Optuna}.

The interpretability of C3NN mainly exists in the following three aspects (Section \ref{sec:model_architecture}): (i) the output moment $c_{\alpha}^{(N)}$ of a specific order $N$ can be mathematically expressed in terms of the correlation function at the same order. This is unlike the often hard to interpret summary statistics extracted by conventional CNN models, (ii) through a regularization path analysis \citep{least_angle_regression_2004, tang2014feature} which is integrated into the classifier part of our model, we can have a quantitative understanding of the relative importance of the different order moments $c_{\alpha}^{(N)}$ in contributing to the model's classification power. Moreover, (iii) we can investigate the trained filter weights by connecting individual pixels to form the configuration of any given NPCF. The filter weights for a given NPCF configuration directly allow us to rank different correlation function configurations within a given moment. 

Since this is the first time we introduce this architecture into the field of cosmology, we focus on the relatively simple task of binary classification and implement C3NN on three tests including both proof of concept as well as application to simulated cosmological data. The results are summarized as follows:
\begin{itemize}
    \item \textit{Gaussian random fields with different correlation lengths} (Section \ref{sec:gaussian_fields})

    The performance of C3NN to distinguish between two Gaussian random fields with different correlation lengths is in line with theoretical expectations. The model successfully achieves $100\%$ classification based on the second moment $c_1^{(2)}$ alone. Its trained filter weights reflect the respective correlation lengths of the two classes of training data. We find good agreement between the distributions of each order of moment measured by C3NN and the corresponding theoretical prediction where we combine the trained filter weights and the analytical calculation of all possible correlation functions at each order.  
    
    \item \textit{Gaussian and Log-normal random fields with the same 2PCF but different higher-order moments} (Section \ref{sec:gaussian_lognormal_fields})

    In this test we apply C3NN to distinguish a Gaussian from a Log-normal random field which start to differ from each other at the 3rd moment. C3NN reaches a perfect classification by first activating and tapping into the information in the 3rd moment of the field $c_\alpha^{(3)}$ and then via the field's first moment $c_1^{(1)}$ (arising due to finite field map sizes). The $c_1^{(3)}$ activation is consistent with the theoretical expectation since the two random fields share the same 2PCF but not the 3PCF. We further demonstrate that the activation of $c_1^{(1)}$ can separate the mapped distribution from the two classes in $c_1^{(1)}-c_1^{(3)}$ plane completely (see lower panel in Figure \ref{fig:path_analysis_grf_lognormal}). Similar to the previous test, we also find good agreement between the the distributions of each order of moment output by C3NN and the corresponding theoretical predictions.
    
    \item \textit{N-body simulated weak lensing convergence fields with $w_0=-1.05$ and $-0.95$} (Section \ref{sec:tomographic_analysis_noiseless} $\&$ \ref{sec:tomographic_analysis_noisy})

    Finally we investigate the classification power of C3NN on weak lensing convergence maps between two simulated dark energy scenarios of $w_0=-1.05$ and $-0.95$. C3NN can maximally reach a classification accuracy around $70\%$ with our training setups for a 400 deg$^2$ survey map. We find that the classification power of the 2nd moment $c_1^{(2)}$ is robust against varying smoothing scales or shape noise but higher-order moments can be significantly impacted. In all our tests, $c_1^{(2)}$ and $c_1^{(3)}$ dominate the classification accuracy and the activation of $c_1^{(3)}$ is the major complementary component to $c_1^{(2)}$ (Figure \ref{fig:ccnn_w0_noiseless_noisy}). On the other hand, higher-order moments of convergence beyond 3rd order in total do not contain sufficient information to classify between the two different dark energy equation of state parameters within the context. This suggests that including more observables besides convergence in the input data and using C3NN (or the conventional 2PCFs and 3PCFs) to measure the cross-correlations among these different observables may be a more efficient approach.
\end{itemize}

Overall, we show that the architecture of C3NN contains novel features which can be robustly and quantitatively interpreted and are rarely seen in the application of machine learning tools to cosmology nowadays. Through multiple tests, we prove its validity and reveal its potential to provide us with physical insights. C3NN can have many interesting extensions such as simulation-based inference with the CNN based N-point moment generator or 3D C3NN which can measure moments in the context of galaxy clustering. All these can offer us promising opportunities for future applications of this architecture in cosmology. 

\vspace{5mm}
\begin{acknowledgments}
We would like to thank Alexandre Barthelemy, Gabriele Cesa, Sandrine Codis, Fulvio Ferlito, Oliver Friedrich, Laila Linke and Tilman Tr$\rm \ddot{o}$ster for very helpful comments
and discussions at various stages of this project. We would also like to give credits to CosmoGridV1 which was created by Janis Fluri, Tomasz Kacprzak, Aurel Schneider, Alexandre Refregier, and Joachim Stadel at the ETH Zurich and the University of Zurich. We acknowledge funding by the Deutsche Forschungsgemeinschaft (DFG, German Research Foundation) under Germany’s Excellence Strategy—EXC-2111—390814868. Some of the  numerical calculations have been carried out on the ORIGINS computing facilities of the
Computational Center for Particle and Astrophysics (C2PAP).
\end{acknowledgments}

\vspace{5mm}
\software{ChainConsumer \citep{chainconsumer},
          ESCNN \citep{escnn_cesa_2022},
          FLASK \citep{xavier2016},
          healpy \citep{Gorski2005,Zonca2019}, 
          Optuna \citep{optuna_2019},
          UFalcon \citep{ufalcon_1, ufalcon_2, ufalcon_3}
          }

\bibliography{sample631}{}
\bibliographystyle{aasjournal}
\end{document}